\documentclass[aps,twocolumn,prd,preprintnumbers,amsmath,amssymb,superscriptaddress,nofootinbib]{revtex4-1}
\pdfoutput=1

\usepackage[export]{adjustbox}

 \usepackage{color}
 \usepackage{epsfig}
 \usepackage{ulem}

\definecolor{White}{rgb}{1,1,1}
\definecolor{Red}{rgb}{1,0.1,0}
\definecolor{LightYellow}{rgb}{1,1,.875}
\definecolor{SteelBlue}{rgb}{.273,.508,.703}
\definecolor{navy}{rgb}{0,0,.5}
\definecolor{LightCyan}{rgb}{.875,1,1}
\definecolor{DarkRed}{rgb}{.543,0,0}
\definecolor{HotPink}{rgb}{1,.41,.70}
\definecolor{ForestGreen}{rgb}{.13,.54,.13}
\definecolor{OliveDrab}{rgb}{.42,.55,.14}
\definecolor{MediumBlue}{rgb}{0,0,.80}
\definecolor{RoyalBlue}{rgb}{.25,.41,.88}
\definecolor{DeepSkyBlue}{rgb}{0,.746,1}
\definecolor{Brown}{rgb}{0.545,0.271,0.074}
\definecolor{Purple}{rgb}{0.637,0.285,0.641}

\def\bea{\begin{eqnarray}}
\def\eea{\end{eqnarray}}
\def\bec{\begin{center}}
\def\ec{\end{center}}

\def\beq{\begin{equation}}
\def\eeq{\end{equation}}

\newcommand\lsim{\mathrel{\rlap{\lower4pt\hbox{\hskip1pt$\sim$}}
    \raise1pt\hbox{$<$}}}
\newcommand\gsim{\mathrel{\rlap{\lower4pt\hbox{\hskip1pt$\sim$}}
    \raise1pt\hbox{$>$}}}
\def\bea{\begin{eqnarray}}
\def\eea{\end{eqnarray}}
\def\ba{\begin{array}}
\def\ea{\end{array}}
\def\bc{\begin{center}}
\def\ec{\end{center}}

\begin{document}

\title{Adiabatic Electroweak Baryogenesis Driven by an Axion-like Particle}

\author{Kwang Sik Jeong} 
\email{ksjeong@pusan.ac.kr}
\affiliation{Department of Physics, Pusan National University, Busan 46241, Korea}
\author{Tae Hyun Jung} 
\email{thjung0720@ibs.re.kr}
\affiliation{Center for Theoretical Physics of the Universe, Institute for Basic Science (IBS), Daejeon, 34126, Korea}
\affiliation{Department of Physics, Florida State University, Tallahassee, FL 32306, USA}
\author{Chang Sub Shin} 
\email{csshin@ibs.re.kr}
\affiliation{Center for Theoretical Physics of the Universe, Institute for Basic Science (IBS), Daejeon, 34126, Korea}
%\date{}
\preprint{CTPU-18-34, PNUTP-18-A12}

\begin{abstract}

An axion-like particle (ALP) offers a new direction in electroweak baryogenesis 
because the periodic nature enables it to trigger a strong first-order phase transition insensitively to the decay constant $f$.
For $f$ much above TeV, the ALP-induced electroweak phase transition is approximately described by adiabatic processes, 
distinguishing our scenario for electroweak baryogenesis from the conventional ones.  
We show that, coupled to the electroweak anomaly, the ALP can naturally realize spontaneous electroweak baryogenesis
to solve the matter-antimatter asymmetry problem
for $f$ in the range between about $10^5$~GeV and $10^7$~GeV.  
In such an ALP window, the $CP$ violation for baryogenesis is totally free 
from the experimental constraints, especially from the recently improved limit on the electron 
electric dipole moment. 
Future searches for ALPs could probe our scenario while revealing the connection between 
electroweak symmetry breaking and baryogenesis.

\end{abstract}

\maketitle

\section{Introduction}

The observed matter-antimatter asymmetry in the universe is one of the pieces of strong evidence for
physics beyond the Standard Model (SM). 
The rapid sphaleron transitions in the symmetric phase provide large violation of baryon number,
indicating that the baryon asymmetry may have been generated at the electroweak (EW) epoch. 
This scenario of electroweak baryogenesis (EWBG) looks 
quite natural and attractive as it  invokes SM baryon number violations and 
is implemented at low temperatures. 
Viable EWBG is achievable in an extension of the SM in which the electroweak phase transition 
(EWPT) is sufficiently strong and $CP$ violation is large during the phase transition.

Recently, the ACME II collaboration improved the limit on the electric dipole moment (EDM) 
of the electron by about one order of magnitude relative to the previous one~\cite{Andreev:2018ayy}.
Although there would still remain an allowed parameter region in the conventional scenarios of EWBG, the improved limit motivates us to consider an orthogonal direction free from the EDM constraints.  
Along this direction, the EDM is no longer a hint for EWBG, and other experimental searches are required to probe the connection between EWPT and baryogenesis. 
For example, if the electron EDM is suppressed in a model for EWBG due to a cancellation among different 
contributions, the associated particles may exhibit a specific pattern of masses and 
couplings~\cite{Bian:2014zka}.
In other models, EWBG may be related with dark matter phenomenology if $CP$ violation for baryogenesis
comes from  a dark sector~\cite{Cline:2017qpe}. 
To avoid EDM constraints, one may instead implement EWBG at a much higher temperature than 
in the conventional scenarios through tachyonic thermal masses~\cite{Baldes:2018nel}, 
for which EWBG yields gravitational waves of much higher frequencies.

%{\color{magenta} For example,  if the electron EDM is suppressed by the cancelation among different contributions, it predicts a specific mass of new particle and couplings~\cite{Bian:2014zka}. If CPV sources relevant for EWBG are coming from a dark sector, it can be related with dark matter phenomenology~\cite{Cline:2017qpe}. 
%Constraints from the EDM can be avoided if EWBG happen at much higher temperatures through negative thermal mass square terms~\cite{Baldes:2018nel}. In such a case, it yields much higher frequencies of the gravitational wave signal.}
    
In Ref.~\cite{Jeong:2018ucz}, we have noticed that the axionic extended Higgs sector
\bea
V = V(|H|^2, \sin(\phi/f), \cos(\phi/f) )
\eea 
provides a simple example of EWBG compatible with the electron EDM bound 
for $f$ above a few TeV if the axion-like particle (ALP) $\phi$ couples to the top quark 
Yukawa operator. 
Here $H$ is the SM Higgs doublet, and $f$ is the ALP decay constant.  
For $f$ above a few TeV,  ALP searches at colliders can give an interesting implication for
the origin of the matter-antimatter asymmetry~\cite{Jaeckel:2015jla,Knapen:2016moh,Mariotti:2017vtv,CidVidal:2018blh}.
In this scenario, $f$ is restricted to be below about $10$~TeV because the bubble wall gets
thicker with $f$, suppressing the charge transport in plasma after scattering off a propagating wall.
In the context of the standard EWBG, a thick wall seems problematic since
baryon asymmetry is mostly produced non-locally through the diffusion of $CP$ asymmetry 
in front of the bubble wall and the $B$-violating sphaleron process active in the symmetric phase
region away from the wall.

In this paper we extend our previous work in Ref.~\cite{Jeong:2018ucz} to explore 
the viability of EWBG at $f$ much above TeV and its connection to ALP searches.
As a source of $CP$ violation, we consider an ALP-dependent EW theta term 
\bea
\label{anomaly} 
\frac{\alpha_W}{4\pi}\Theta_{\rm EW} W^{a\mu\nu}\tilde W^a_{\mu\nu},
\eea
with
\bea
\Theta_{\rm EW} = \frac{\phi}{f}.
\eea
The above coupling can be induced easily, for instance, through loops of extra heavy 
leptons charged under the ALP shift symmetry, $\phi \to \phi +(\rm constant)$.
It turns out that the standard non-local production of baryon asymmetry is highly suppressed, 
but instead sizable baryon asymmetry can be generated locally as a result of $B$ and $CP$-violating processes occurring simultaneously near and across the bubble wall. 
This way, the ALP implements so-called local {\it spontaneous} EWBG.

During EWPT, the ALP field changes its value as $\Delta \phi ={\cal O}(f)$, and thus
the time derivative of $\Theta_{\rm EW}$ acts as a source for the chemical potential 
of the Chern-Simons (CS) number at a given spatial point. 
This leads to the generation of baryon number through the EW anomaly:
\bea
\label{baryogenesis}
\frac{d n_B}{dt} \approx \frac{N_g}{2}\frac{\Gamma_{\rm sph}}{T} 
\frac{d\Theta_{\rm EW}}{dt}  - \Gamma_B n_B, 
\eea
with $N_g=3$ being the number of generations. 
Here $\Gamma_{\rm sph}$ is the sphaleron transition rate  per unit volume,
and $\Gamma_B=  (13 N_g/4) \Gamma_{\rm sph}/T^3$ is the rate of 
the sphaleron-induced relaxation of baryon asymmetry~\cite{Bochkarev:1987wf,Cline:2000nw}.

Local spontaneous EWBG has been studied intensively in the early stage of the development
of EWBG~\cite{Cohen:1991iu,Giudice:1993bb,Dine:1994vf,Joyce:1994bk,Cohen:1994ss,Joyce:1994fu}.
However, it was noted that the $CP$-odd scalar in a two-Higgs doublet model cannot
give sufficient $CP$ violation for baryogenesis without diffusion 
effects~\cite{Dine:1994vf,Joyce:1994bk,Cohen:1994ss}. 
At that time, there was also a large uncertainty in the baryon asymmetry estimation 
due to the lack of numerical understanding of how $\Gamma_{\rm sph}$ changes 
with the Higgs vacuum expectation value.
Furthermore, the realistic bubble wall is not so thick in the usual EWBG models, for which the 
out-of-equilibrium process and charge transport are quite important and most of the baryon 
asymmetry is produced ahead of the bubble wall. 
The situation is quite different for EWPT triggered by the ALP
because the bubble wall width is much larger than the diffusion length scale in thermal bath. 
This implies that baryogenesis occurs in the adiabatic limit. 
The recent lattice calculation of the sphaleron rate shows the dependence 
on temperature and the Higgs vacuum expectation value~\cite{DOnofrio:2014rug}.

On one hand, an EW theta term varying during EWPT has been studied before,
for instance, see Refs.~\cite{Shaposhnikov:1987pf,Dine:1990fj},
but mostly in the context of cold baryogenesis~\cite{GarciaBellido:1999sv}.
Those
models rely on efficient production of Higgs winding numbers, which could be achieved
through a preheating stage with an inflaton coupled the Higgs sector~\cite{GarciaBellido:1999sv}, 
or a delayed first-order phase transition induced by conformal symmetry breaking and 
subsequent bubble collisions~\cite{Konstandin:2011ds,Servant:2014bla}. 
Such a violent environment can generate unstable Higgs winding numbers which are large
enough to decay through the production of $CP$-violating CS numbers. 
Another way to induce a time-dependent EW theta term is to consider $CP$ violation 
from an axion anomalously coupled to a confining hidden gauge group, and its transmission to 
the SM via messengers~\cite{Craig:2010au}.
Then, assuming some mechanism for a strong first-order EWPT, EWBG would be realized
in the parameter space where the axion 
slowly rolls and the messenger masses significantly change during EWPT.

%{\color{magenta}  
%Apart from a condition of successful EWPT, 
%a time dependent EW theta term can be generated by various ways such as \cite{Craig:2010au}, in which  
%thermal dynamics of the confining hidden sector with a slowly rolling ALP play the important role during EWPT. Since the strong first order phase transition is just assumed in \cite{Craig:2010au},  there is no further connection between the ALP and the strength of EWPT.

%In our scenario, there is no violent out-of-equilibrium process, and all stages of baryogenesis are nearly smooth, and the ALP plays the key role.  
%This also allows to make a concrete prediction for the baryon asymmetry while establishing 
%an interesting and meaningful relation between EWBG and ALP searches.} 

In our scenario, the ALP plays the essential role in both EWPT and baryogenesis.
We also note that there is no violent out-of-equilibrium process, and all stages of baryogenesis 
proceed nearly smoothly. 
This allows us to make a concrete prediction for the baryon asymmetry while establishing
an interesting and meaningful relation between EWBG and ALP searches. 
We find that, feebly coupled to the Higgs sector and EW anomaly, the ALP can naturally solve 
the puzzle of the matter-antimatter asymmetry in the universe.
Successful baryogenesis is achieved for $f$ below $10^8$~GeV, and the model is totally 
free from the EDM constraints for $f$ much above TeV.  
The viable window is $f$ between about $10^5$ and $10^7$~GeV, 
or equivalently the ALP mass roughly equal to $m^2_W/f$, i.e.~in the MeV to GeV range,
%ALP mass {\color{blue} of ${\cal O}(m_W^2/f)$},
%in the MeV to GeV scale, 
once the constraints on ALP-Higgs mixing from 
various experiments are imposed.  
Our scenario therefore encourages experimental searches for ALPs in the indicated window
of parameter space, which would otherwise fall short of strong theoretical interest.

This paper is organized as follows. 
In Sec.~\ref{sec:EWPT}, we show that a strong first-order phase transition is achievable
in the Higgs potential modified by the ALP even in the weakly coupled regime with $f$
much above TeV, and then discuss essential features of the ALP-induced EWPT.
In Sec.~\ref{sec:BG}, we examine spontaneous EWBG naturally realized by the ALP
via its coupling to the EW anomaly. 
The ALP is subject to various experimental constraints because it mixes with the Higgs boson.
We summarize the constraints on the ALP properties in Sec.~\ref{sec:EC}.  
Sec.~\ref{sec:Con} is devoted to the conclusions.

\section{Electroweak Phase Transition}
\label{sec:EWPT}

In this section we discuss how a strong first-order phase transition is achieved in the Higgs 
potential modified by the ALP.
For an explicit model, we consider the case in which 
the ALP $\phi$ couples to the mass squared operator of the Higgs field $H$
%the ALP {\color{blue} ($\phi$)} couples to the Higgs {\color{blue} ($H$)} mass squared
%operator
\bea
V = \lambda |H|^4
+ \mu^2_H (\theta) |H|^2 
+ V_0(\theta) + \Delta V_{\rm TH},
\label{potential}
\eea
for $\theta\equiv \phi/f$, with 
\bea 
\mu^2_H(\theta)  &=& \mu^2 - M^2 \cos(\theta+ \alpha),
\nonumber \\
V_0(\theta) &=& -\Lambda^4 \cos\theta + {\rm constant},
\eea  
under the assumption that  $f$ is above the EW scale while
other mass parameters $\Lambda$, $\mu$ and $M$
%other involved mass parameters {\color{blue} $\Lambda$, $\mu$, $M$} 
are around or below the EW scale.
Here $\alpha$ is a constant phase, and $\Delta V_{\rm TH}$ includes thermal corrections.
It is worth noticing that the ALP-dependent terms are generated in a controllable way
if the ALP shift symmetry is broken solely by nonperturbative 
effects~\cite{Jeong:2018ucz, Graham:2015cka}.

At a temperature much below $f$ but around or above the EW scale, 
thermal corrections to $V$ from the SM plasma are still sizable 
while those due to the ALP interactions are suppressed by powers
of $T/f$. 
This implies 
\bea
\Delta V_{\rm TH} \simeq \Delta V^{\rm SM}_{\rm TH}(|H|^2),
\eea
where $\Delta V^{\rm SM}_{\rm TH}$ includes thermal corrections only
from the SM particles.
The thermal evolution of the scalar fields is thus described as follows. 
In phase transition, the most important role is played by the contribution of 
$\Delta V^{\rm SM}_{\rm TH}$ to the Higgs quadratic term.
The thermal corrected Higgs mass squared is approximately given by
\bea
\mu^2_{HT}(\theta) \simeq
\mu^2 - M^2\cos(\theta + \alpha ) + c_h T^2,
\eea
for a positive coefficient $c_h$ determined by SM couplings.
For sufficiently high temperatures, $\mu^2_{HT}$ is positive
for all values of $\theta$, making $V$ develop a unique minimum
at $(\theta,H)=(0,0)$.
For $M^2 > \mu^2$,  it is clear that $\mu^2_{HT}$ becomes
negative in a certain range of $\theta$ if the temperature drops
sufficiently, implying that there appears an additional local minimum 
at $\theta \neq 0$ and $H\neq 0$. 
The two minima are degenerate when the universe cools down to
$T=T_c$, and then a phase transition happens from the symmetric phase
to the broken one at a temperature below $T_c$. 
After the phase transition, the scalar fields roll toward the true vacuum.

For the scalar potential (\ref{potential}),  $\phi$ and $h$ form two mass eigenstates $\varphi_L$ and $\varphi_H$ with 
temperature-dependent
masses $m_L$ and $m_H$, respectively,
where $h=\sqrt2 |H^0|$ denotes the neutral Higgs scalar. 
For $f$ much above the EW scale, the light scalar $\varphi_L$ is mostly the ALP and has a mass,
$m_L  \sim m^2_H/f$.
As can be deduced from such a large mass hierarchy, the field evolution occurs mainly along 
the direction of the light ALP-like field, and the fluctuation along the direction of the heavy Higgs-like field
is quickly damped within the time scale of order $1/m_H$. 
This feature has been explicitly shown in the appendices~\ref{appendix:bounce} and~\ref{appendix:res_osc}.
One can thus examine the structure of phase transition within an effective theory 
constructed by integrating out the heavy Higgs field via the equation of motion 
\bea
\left.\frac{\partial V}{\partial h} \right|_{h=\hat h(\phi)} = 0,
\eea
where the solution $\hat h$ is found to be
\bea\label{eq:hhat}
\hat h(\phi) \simeq
\left\{
  \begin{array}{ll r}
    0 & \qquad{\rm for} &  \mu^2_{HT}(\phi)\geq 0   \\
  \sqrt{ -\frac{1}{\lambda} \mu^2_{HT}(\phi) } & \qquad{\rm for} & \mu^2_{HT}(\phi)<0
  \end{array}
  \right..
\eea   
We note that a more precise solution is obtained if one includes contributions from
$\Delta V^{\rm SM}_{\rm TH}$ to the Higgs cubic and quartic terms.
At a temperature at which $V$ develops two minima, the effect of such contributions is to make 
the EW minimum deeper and farther from $h=0$, because thermal corrections are Boltzmann-suppressed 
at Higgs field values larger than $T$.
Therefore, with the precise solution, one would find that the suppression of sphaleron
processes in the broken phase is strengthened as preferred for EWBG. 
Keeping this in mind, we take Eq.~(\ref{eq:hhat}) as a good approximation.
%because it allows to find almost whole region of the parameter space leading to a strong first-order 
%phase transition and to estimate baryon asymmetry precisely enough without changing 
%the qualitative features of EWBG.  

%Actually, in order to evaluate correct values of $\hat h(\phi)$ for  $\hat h(\phi) \gtrsim T$, 
%we also have to add cubic and quartic potentials of $\Delta_{\rm SM} V_T(|H|)$. 
%It is easy to see that the full contributions give a deeper minimum of $V$  at 
%a larger value of $\hat h$ compared to (\ref{eq:hhat}), because thermal corrections are Boltzmann suppressed as $\hat h$ is greater than $T$. Therefore 
%in the broken phase with   precise values of $\hat h(\phi)$, the sphaleron process is more  suppressed. In this sense, we can take (\ref{eq:hhat}) as the good approximation  without worrying about additional washout effect for baryogenesis. 

It is straightforward to see that the effective potential of the light field reads
\bea
V_{\rm eff}(\phi) = \Lambda^4 (\cos\theta_0 - \cos\theta)
-\frac{\lambda}{4} \left( \hat h^4(\phi)-v^4_0 \right),
\eea 
where the true minimum at $T=0$ is located at 
$(\theta,h)=(\theta_0,v_0)$. 
Fig.~1 in Ref.~\cite{Jeong:2018ucz} illustrates how $V_{\rm eff}(\phi) $ changes with $T$ and 
how it is projected from the full potential $V(H, \phi)$. 
In what follows, we will parametrize the potential in terms of three dimensionless parameters
\bea
\quad \alpha, 
\quad
\epsilon \equiv  
\frac{\sqrt{2\lambda} \Lambda^2}{M^2},
\quad
r \equiv \frac{\sqrt{2} \Lambda^2}{\sqrt\lambda v^2_0},  
\eea
by imposing the condition $v_0=246$~GeV
and the observed Higgs boson mass to fix $\lambda$ and $\mu$. 
From the scalar potential, one finds   
\bea\label{sintheta0}
\sin\theta_0 = \frac{-\sin\alpha}{\sqrt{1+2r\epsilon \cos\alpha + r^2 \epsilon^2} },
\eea
where
the overall sign of $\cos\theta_0$ is fixed by the minimization condition.

Let us briefly illustrate the procedure of a first-order phase transition driven by the ALP.  
At high temperatures, the minimum of $V_{\rm eff}$ is located at $\theta=0$
because large thermal corrections lead to $\hat h(\theta)=0$ in the whole range of 
$\theta$. 
The initial position of the ALP at a high temperature is generally displaced from the potential
minimum $\theta=0$, but its effect on the phase transition can be safely ignored as long as 
the ALP potential is developed at a temperature much above the weak scale. 
For instance, if generated by hidden QCD~\cite{Jeong:2018ucz}, 
the ALP mass grows according to 
\bea
m_\phi(T) \propto \left(\frac{\Lambda_c}{T} \right)^\ell,
\eea
as the universe cools down for $T> \Lambda_c$, and it reaches the value equal to 
$\Lambda^2/f$ at $T$ around the hidden confinement scale $\Lambda_c$.
Here, $\ell =( 11N_c + N_f)/6-2$ and $\Lambda^4 = m_N \Lambda^3_c$
for a confining SU$(N_c)$ with $N_f$ vector-like quarks having masses 
$m_N \ll \Lambda_c$.
The ALP starts coherent oscillations at $T=T_{\rm osc}$ when the Hubble expansion 
rate becomes comparable to the ALP mass.
Note that $T_{\rm osc}$ can be well above the weak scale because
$\Lambda_c$ is higher than $\Lambda$. 
In the case with $N_c=3$ and $N_f=1$, oscillation starts at 
\bea
T_{\rm osc} \simeq 25\,{\rm TeV}
\left(\frac{\Lambda^2/f}{10{\rm MeV}}\right)^{3/17}
\left(\frac{\Lambda_c}{{\rm TeV}}\right)^{11/17},
\eea
and its amplitude is given by 
\bea
\theta_{\rm osc}(T) &\simeq&
\theta_{\rm ini}
\left(\frac{m_\phi(T_{\rm osc})}{m_\phi(T)}\right)^{1/2} 
\left(\frac{T}{T_{\rm osc}}\right)^{3/2}
\nonumber  \\
&=&
2.7\times 10^{-6}\,
\theta_{\rm ini}
\left(\frac{T}{\Lambda_c}\right)^{3/2}
\left(\frac{\Lambda_c}{{\rm TeV}}\right)^{20/17},
\eea
since the ALP number density scales as $T^3$ during coherent oscillations.
Here $\theta_{\rm ini}$ is the initial misalignment angle of the ALP. 
The above shows that $\theta_{\rm osc}$ becomes negligibly small
at the time of EWPT if the hidden QCD confines at a scale above 
the weak scale. 

When the universe sufficiently cools down, there appears a region of 
$\theta$ with nonvanishing $\hat h$, which is around $\theta=\epsilon-\alpha$. 
For $\alpha\neq 0$, $V_{\rm eff}$ develops two degenerate minima separated by
a barrier at the critical temperature $T=T_c$.
Then, EW bubbles of the broken phase are nucleated and expand. 
The EW minimum gets deeper than the symmetric one as $T$ decreases, and the bubble
nucleation rate per unit volume exceeds the Hubble expansion rate at $T=T_n$. 
Finally the potential barrier between two minima disappears at $T=T_2$: 
\bea 
T_2 = \sqrt{
1  - \frac{r}{\epsilon} \Big(
\cos(\theta_0 + \alpha) - \cos\alpha \Big)
}\, T^{\rm SM}_c ,
\eea
where $T^{\rm SM}_c = \sqrt{\lambda/c_h}\,v_0\simeq 150$~GeV is the critical temperature
for the SM Higgs sector.
The phase transition is thus first-order, and is strong if $v_c/T_c >1$ with 
 $v_c$ being the Higgs vacuum expectation value at $T_c$. 
Interestingly, a strong first-order phase transition is achievable even for $f$ much above
the EW scale, i.e.~in the weakly coupled limit, which distinguishes our model from 
the conventional approaches.

\begin{figure}[t] 
 \begin{center}
 	\includegraphics[width=0.37\textwidth]{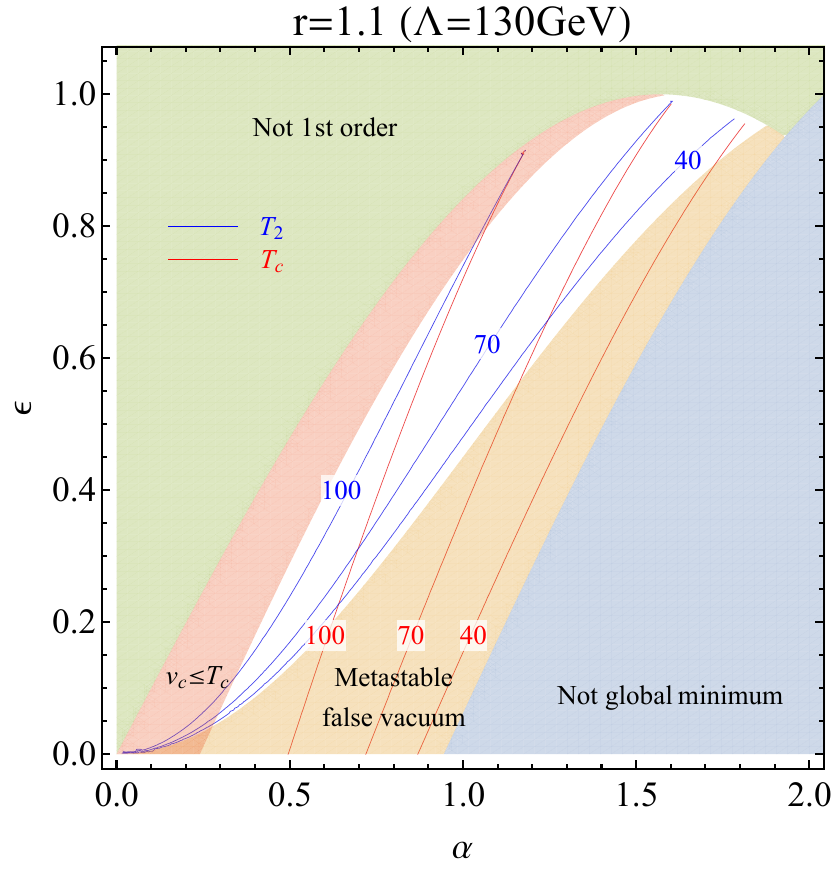} 
 \end{center}
 \caption{
EWPT in the Higgs sector modified by the ALP for $r=1.1$.
A strong first-order phase transition occurs in the white region, insensitively to 
the value of $f$.
The red and blue lines show the critical and bubble disappearing temperatures
in the GeV unit, respectively. 
 }
\label{fig:T2Tc}
\end{figure}

Fig.~\ref{fig:T2Tc} illustrates how EWPT takes place depending on $\alpha$ and $\epsilon$
for the case with $\Lambda=130$~GeV, which corresponds to $r\simeq 1.1$.  
A first-order phase transition is achieved in the white and red regions, and it is strong
in the white region.
In the region of a first-order phase transition, we also show the constant contours 
of $T_2$ by blue lines, and those of $T_c$ by red lines. 
The blue region leads to an EW minimum higher than the symmetric minimum, while
the orange region is excluded because the vacuum transition rate to the EW minimum
is highly suppressed for $f$ above TeV due to a barrier remaining at $T=0$.
In the green region, the phase transition is not first-order as in the SM. 
We note that the indicated lines and regions in the figure change only slightly with $f$ because 
the potential for fixed $\theta$ does not rely on it at the tree level.
Let us shortly discuss the generic behavior for a different value of $r \propto \Lambda^2$.
If one increases $r$, a strong first-order phase 
transition occurs at smaller values of $\epsilon$ for a given $\alpha$ because it requires
sizable $\sin\theta_0$. 
Thus, the white region in Fig.~\ref{fig:T2Tc} moves to the bottom right.
In the opposite case with a smaller value of $r$, it moves to the top left.
The remaining qualitative behaviors are the same as before. 
%{\color{blue} 
%Let us shortly discuss about the generic behavior 
%when we take different $\Lambda$. 
%If we increase $\Lambda$, $r$ also increases. Therefore 
%in order to get a sizable $\sin\theta_0$ of Eq.~(\ref{sintheta0}), which is 
%relevant for a strong first-order PT condition, 
%the allowed values of $\epsilon$ will decrease for a given $\alpha$, 
%and the allowed value of $\alpha$ will increase for a given $\epsilon$. 
%The white region of Fig.~\ref{fig:T2Tc} will move to the bottom right. 
%If $\Lambda$ is decreasing, then it will move to the top left.  
%The remaining qualitative behavior is the same as Fig.~\ref{fig:T2Tc}}

Let us examine the phase transition in more detail. 
The bubble nucleation rate is given by $T^4 e^{-S_3/T}$,
where $S_3$ is the Euclidean action of an $O(3)$ symmetric critical bubble.
For $f$ above TeV, the contribution to $S_3$ from Higgs kinetic terms
is highly suppressed, and tunneling occurs dominantly along the ALP direction.
Interestingly, 
combined with the insensitivity of the scalar potential to $f$ for given $\theta$,
this leads to the approximate scaling laws
\bea
S_3 &\propto& f^3,
\nonumber \\
R_c &\propto& f,
\eea 
where $R_c$ is the radius of the critical bubble. 
See the appendix \ref{appendix:bounce} for the details.  
At temperatures around $T_n$, one can thus take an approximation 
\bea \label{eq:S3}
\frac{S_3}{T} \propto (T-T_2)^n f^3,
\eea
for a positive constant $n$ of order unity, where we have used that 
$S_3=0$ at $T=T_2$ because there is no potential barrier. 
Thus there are characteristic features specific to our scenario.
One is that the bubble nucleation temperature, which is determined by $S_3/T\approx 130$,  
is close to the barrier disappearing temperature
\bea
T_n \sim T_2,
\eea
where the difference between the two is suppressed by a factor of $f^{-3/n}$. 
Another distinctive feature is that bubbles are formed with a thick wall roughly proportional
to $f$, and the phase transition proceeds rather smoothly with nucleation of bubbles.
This implies that the phase transition is approximately adiabatic during baryogenesis, 
and diffusion through the bubble wall is not efficient
for large $f$.

It follows from the scaling behavior of $S_3$ that the duration of phase transition  
decreases with $f$ as  
\bea
\Delta t_{\rm PT} \simeq
\frac{6}{-d(S_3/T)/dt|_{T_n}}
\propto 
\frac{1}{f^{3/n}}.
\eea 
For $r$ of order unity, a numerical analysis shows
\bea
\Delta t_{\rm PT} \sim \frac{10^{-2}}{H}
\left( \frac{1{\rm TeV}}{f} \right)^{3/n}, 
\eea 
with $1\lesssim n \lesssim 2$, in the parameter region of a first-order phase transition. 
Here $H$ is the Hubble expansion rate at $T=T_n$.

On the other hand, the wall width in the rest frame of a bubble wall can be regarded
as the size of the critical bubble, and it is given by  
\bea
L_w \sim \sqrt{\frac{\Delta \Phi_c^2}{\Delta V_c}} 
\sim \frac{f}{\Lambda^2},
\label{rc}
\eea 
where $\Delta \Phi_c$ is the field variation during tunneling, and 
$\Delta V_c \sim \Lambda^4 (\Delta \Phi_c/f)^2$ is the height 
of the potential barrier.
One can see that the bubble wall is thick, $L_w\gsim 100/T_n$, for $f$ above $10^4$~GeV
and $T_n$ around $50$~GeV.  
This corresponds to the adiabatic regime, in which non-thermal enhancement of baryon production
is expected
neither from particle diffusion~\cite{Joyce:1994bk,Cohen:1994ss} nor 
the classical dynamics of fast Higgs quenching~\cite{Konstandin:2011ds}.

If $f$ is even larger to give $L_w > v_w \Delta t_{\rm PT}$ 
with $v_w$ being the wall velocity, the phase transition 
proceeds via bubble nucleation but without substantial expansion of bubbles.
This happens when $f\gtrsim 10^6$~GeV for $n=1$,
and $f\gtrsim 10^8$~GeV for $n=2$, where we have taken $v_w\sim 0.1$.
It is also important to note that bubble nucleation is followed by rolling of the ALP toward 
the true minimum of the potential after tunneling.
The phase transition looks smooth for $T_n$ close to $T_2$, but it is definitely distinguishable 
from a second-order one because its large mass makes the ALP evolve
much more quickly compared to the cooling rate of the universe.

For a final remark in this section, we note that a singlet scalar can play a similar role in EWPT
as the ALP in our model under certain conditions on its couplings.  
Let us consider an extension with a real scalar $s$:
\bea
V = \lambda |H|^4 +  \mu^2_H(s) |H|^2 + V_0(s).
\eea 
For the scalar feebly coupled to $H$, 
a first-order phase transition is still possible if $\mu^2_H$ is negative in a finite range of $s$, 
and $V_0$ is bounded from below and has a single minimum lying in the region where $\mu^2_H$ 
is negative.
Here $V_0$ should not be too steep around the minimum so that the high temperature potential 
can properly develop symmetric and EW minima separated by a barrier.
For a simple example, we consider  
\bea
\mu^2_H &=& -\mu^2 + \lambda_{hs}  (s - \mu_s)^2,
\nonumber \\
V_0 &=&  \sum^{4}_{n=1} \lambda_n  \mu^{4-n}_s s^n,
\eea
for a positive coupling $\lambda_{hs}\ll 1$. 
Under the assumption for simplicity that $V_0$ has a single minimum at $s=0$,
the conditions for a first-order phase transition read  
\bea
&&  
0< -\mu^2_H(s=0) \sim v^2_0, 
\nonumber \\
&&
V_0(s=\mu_s) - V_0(s=0) < \frac{\mu^4}{4\lambda} \sim v^4_0,
\eea
implying   
$\lambda_{hs} \sim (v_0/\mu_s)^2$ and 
$\sum_n \lambda_n \lesssim (v_0/\mu_s)^4$. 
Such a hierarchical structure of singlet couplings would indicate some underlying symmetry.   
An ALP is therefore a natural candidate because its couplings are controlled by the associated 
shift symmetry. 
In this case, $\mu_s$ corresponds to the ALP decay constant. 
An important feature of the ALP extension is that one can control separately the strength of
couplings and the strength of EWPT since the latter is insensitive to the decay constant.
Furthermore, the periodic nature allows us to avoid the instability problem of the scalar potential
independently of the details of the model.

\section{Baryogenesis} 
\label{sec:BG}

Coupled to the Higgs mass squared operator, the ALP makes EWPT strongly first-order
in a wide range of parameter space including the weakly coupled regime with large $f$.
Furthermore, its coupling to the EW anomaly provides a sizable chemical potential 
for the CS number during phase transition.
As a result, the ALP naturally realizes spontaneous EWBG to solve the matter-antimatter 
asymmetry problem.

A distinctive feature of ALP-induced EWPT is that it is approximately adiabatic 
for $f$ above $10$~TeV. 
Then, a thick bubble wall makes diffusion effects inefficient, implying that non-local baryon 
production can be neglected for large $f$, where the wall gets thicker proportional to $f$ 
as discussed in Sec.~\ref{sec:EWPT}. 
The ALP induces local baryon production by providing a CS chemical potential. 
Another intriguing feature, which will be discussed below, is that baryogenesis proceeds 
almost isothermally if $T_n$ is above about $30$~GeV.
This makes the situation simple to analyze.

\begin{figure*}[t]
 	\includegraphics[height=0.22\textheight]{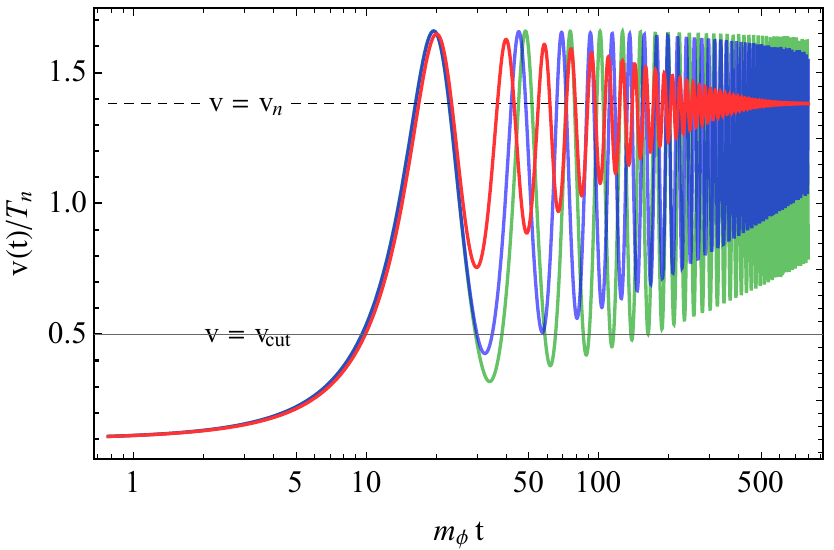}
    \hspace{0.4cm}
 	\includegraphics[height=0.22\textheight]{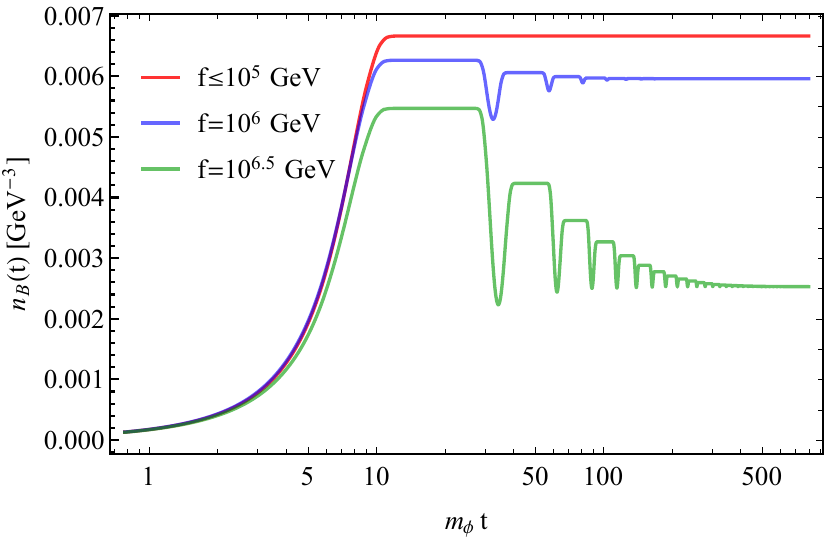}
 	\caption{
    Evolutions of the Higgs background field value (left) and baryon number density (right) 
    in the case with $\alpha=1.4$, $\epsilon=0.95$ and $r=1.1$ for different values of $f$
    as indicated in the figure. 
    Here $m_\phi$ is the ALP mass. 
    The Higgs background field value oscillates about the potential minimum $v=v_n$, and sphalerons
    are active for $v$ smaller than $v_{\rm cut}$.  
    The right panel shows that baryon asymmetry is efficiently produced during the first falling toward $v=v_n$,
    and then it is washed out by later oscillations if the oscillation reaches the region with $v\lesssim v_{\rm cut}$.
     }
 \label{fig:evolution}
 \end{figure*}

Let us now examine the ALP evolution during phase transition.   
The ALP undergoes an underdamped oscillation inside bubbles
following the equation of motion  
\bea 
\label{scalar_evolution}
\ddot \phi + \left(3H +   \Upsilon_\phi^{\rm eff}   \right) \dot \phi  
+ \frac{dV_{\rm eff}}{d\phi} = 0,
\eea
where 
the dot denotes a derivative with respect to time $t$, and 
$\Upsilon_\phi^{\rm eff}$ is the effective energy transfer rate from the ALP field 
to other particles and bubbles.   
For $f$ above about $10$~TeV, the typical time scale characterizing the dynamics of
SM thermal plasma,
which is roughly $1/(\alpha^2_s T_n)$, is much shorter than the time scale 
of the field variations approximately $1/m_\phi$.  
The baryon asymmetry can then be numerically calculated by solving Eqs.~(\ref{baryogenesis}) 
and (\ref{scalar_evolution}) as shown in Fig.~\ref{fig:evolution}.

On the other hand, it is also possible to analytically understand how baryogenesis proceeds.
The solution to Eq.~(\ref{baryogenesis}) can be written in the integral form
\bea 
n_B(t) = \int_0^t 
dt' \frac{3\Gamma_{\rm sph} }{2T} \dot\theta\,
{\rm Exp}\left[ -\int^t_{t'} dt''\frac{39\Gamma_{\rm sph}}{4T^3}\right].
\eea 
%
%{\color{blue} 
%Here we assume that a typical time scale 
%characterizing a dynamics of the thermal plasma ($\sim 1/\alpha_s^2 T_n$) is much shorter than that of the field  variations ($\sim 1/m_\phi$). 
%This is valid for $f$ above about $10$~TeV. 
%The baryon asymmetry can be calculated numerically by solving Eq.~(\ref{baryogenesis}) and Eq.~(\ref{scalar_evolution})  as Fig.~\ref{fig:evolution}.
%On one hand, there is the analytic way to understand how baryogenesis proceeds. 
%First of all, the solution to Eq.~(\ref{baryogenesis}) can be written as 
%the integral form:
%\bea 
%n_B(t) =\int_0^t 
%dt' \frac{3\Gamma_{\rm sph}(t)}{2T}\frac{d\theta}{dt'} 
%\exp\Big[- \int_0^{t'} dt''\frac{39\Gamma_{\rm sph}(t)}{4T^3}\Big].
%\nonumber\\
%\eea  
It is convenient to separate the ALP evolution 
into two parts, the first falling toward the potential minimum and later oscillations.
It is during the first falling that baryon asymmetry is efficiently created while passing 
the region with small $v$ where sphalerons are active.  
The relaxation of baryon asymmetry is negligible at this stage.
On the contrary, the effect of later oscillations is only to wash out the baryon asymmetry  
because a cancellation occurs between baryon and anti-baryon numbers produced by 
the CS chemical potential at each oscillation. 
Using the fact that the first falling and later oscillations of the ALP play different roles in baryogenesis, 
one can reduce the solution of the integral form to\footnote{
More generally, the ALP coupling to the EW anomaly reads
$
\Theta_{\rm EW} = N_{\rm EW} \phi/f,
$
where an integer $N_{\rm EW}$ is model-dependent.
We shall take $N_{\rm EW}=1$ throughout the paper, but one can consider a different value to enhance
the baryon asymmetry.  
}
\bea
\label{nB}
n_B = n_0 e^{-K_\phi},
\eea
in which $n_0$ is determined by the baryon asymmetry produced during the first falling
\bea
n_0 \simeq 27\alpha^5_W T^3_n \Delta\theta, 
\eea
%\bea
%n_B \simeq 
%\left(27\alpha^5_W T^3_n \Delta\theta \right) e^{-K_\phi},  
%\label{nB}
%\eea
and the exponential factor represents the washout during oscillations 
\bea 
K_\phi \simeq
\frac{351\alpha_W^5 T_n}{2} 
\left( \Delta t_0 
+ 2\sum^{N_{\rm osc}}_{i=1} \Delta t_i \right).
 %
% 
%\left(\int_0^{\Delta t_0}+  \sum^{N_{\rm osc}}_{\ell=1}
%\int^{t_\ell + \Delta t_\ell}_{t_\ell -\Delta t_\ell}\right)\frac{351\alpha_W^5 T_n}{2}.
\eea 
Here $\Delta t_0$ is the duration of the first falling, and
$N_{\rm osc}$ counts the number of oscillations such that
sphalerons are unsuppressed during $t_i -\Delta t_i < t < t_i + \Delta t_i$ 
around the peak of the $i$th oscillation. 
The interval of $\theta$ where $\hat h(\theta)$ is 
smaller than $v_{\rm cut}$ during the first falling is estimated to be 
\bea
\Delta\theta
\simeq
\frac{\epsilon}{r \sin\alpha} 
\frac{v^2_{\rm cut}}{v^2_0}, 
\label{deltatheta}
\eea
where we have used that the phaleron rate reads 
$\Gamma_{\rm sph} \approx 18 \alpha^5_W T^4$ if the Higgs background field value is smaller 
than $v_{\rm cut}\simeq 0.5 T$,  and it is exponentially suppressed
otherwise~\cite{DOnofrio:2014rug}.

The correct baryon asymmetry is obtained if the ALP evolution does not cause strong 
washout. 
Let us examine the conditions for this. 
The temperature is kept near $T_n$ during all stages of baryogenesis, and thus 
one needs 
\bea
\frac{v_n}{T_n} \gtrsim 1,
\eea
which is slightly weaker than the condition for a strong first-order phase transition illustrated
in  Fig.\,\ref{fig:T2Tc}. 
Here, $v_n$ is the Higgs vacuum expectation value at $T_n$. 
In addition, the friction term should quickly reduce 
the ALP oscillation amplitude so that the Higgs background field value
\bea
v(t) \equiv \hat h(\theta(t))
\eea 
is smaller than $v_{\rm cut}$ afterward. 
This requires 
\bea
f  < 10^8\,{\rm GeV},
\eea
because the ALP dissipates energy into the background plasma through interactions
with SM particles induced by scalar mixing. 
For $f$ in the opposite region, bubbles do not expand substantially during the phase transition, and so the friction comes dominantly from thermal dissipation from the coupling to the top quarks through the ALP-Higgs 
mixing~\cite{  Wang:1999mb,Mukaida:2012qn}, 
\bea 
\Upsilon_\phi^{\rm eff} \sim 
\left\{
  \begin{array}{ll r}
    y^2_t \Gamma_{\rm th} \sin^2\delta(t)   &  {\rm for} &    y_t v(t) \lesssim \Gamma_{\rm th} \\
    y^2_t \frac{(y_t v(t))^2}{\Gamma_{\rm th}} \sin^2\delta(t)  &  {\rm for}   &  
    y_t v(t) \gtrsim \Gamma_{\rm th}
  \end{array}
  \right.,
\eea
with the thermal width $\Gamma_{\rm th}\simeq 0.1 T$ determined by the top quark Yukawa coupling
$y_t$ and the QCD gauge coupling.  
Here the ALP-Higgs mixing angle $\delta$ changes with time as $v(t)$ does. 
%
%{\color{red}  
%\bea
%\Upsilon_\phi^{\rm eff} &\sim &
% y_t^2 \alpha_{t, g} T \sin^2\delta(t)\quad\ \, {\rm for} \quad
% y_t h \lesssim \alpha_{t,g} T \nonumber\\
% &\sim & y_t^2 \frac{(y_t h)^2}{\alpha_{t, g} T}   \sin^2\delta(t)
% \quad {\rm for} \quad y_t h \gtrsim \alpha_{t, g} T 
%\eea
%where $y_t$ is the top Yukawa coupling and 
%$\alpha_{t, g}= y_t^2/4\pi, g_3^2/4\pi \simeq 0.1$, 
%and the ALP-Higgs mixing angle $\delta$ changes with time 
%as $v(t)$ does. }
See the appendix~\ref{appendix:res_osc} for more discussion on the evolution of Higgs and ALP fields 
including other sources of dissipation. 
In the numerical analysis, we neglect the contribution of ${\cal O}(g_i^2)$ in the
denominator. During the time when the ALP passes the region with $v$ larger than $T_n$, the top quark 
decouples from thermal equilibrium, and dissipation gets suppressed by the Yukawa couplings 
of other light fermions. 
For $f$ around and above $10^8$~GeV, the exponent $K_\phi$ is larger than order unity 
and scales roughly with $1/f^2$, and thus it corresponds to the strong washout regime. 
Note that the ALP decay is highly suppressed, but occurs well before nucleosynthesis 
for $f$ below $10^8$~GeV.

The estimation of baryon asymmetry also requires knowing how much temperature changes
during baryogenesis. 
After the phase transition, the ALP settles down to the potential minimum, and the universe 
heats up to the temperature $T=T_{\rm reh}$ with $T_{\rm reh}$ determined by
\bea
\hspace{-0.3cm}
\left(\frac{T_{\rm reh}}{T_n}\right)^4 \simeq
1 + 0.1 \Big(\frac{\Delta V}{(80\,{\rm GeV})^4}\Big) 
\left(\frac{60\,{\rm GeV}}{T_n}\right)^{4},
\eea
taking $g_\ast=100$.
Here $\Delta V$ is the difference of vacuum energy densities 
of the symmetric and broken phases at $T_{\rm reh}$, and so it is 
a fraction of the former, $V_{\rm eff}(\theta=0)$. 
The above relation indicates that extra entropy production is negligible, i.e.
\bea
T_{\rm reh}\simeq T_n,
\eea
unless $T_n$ is lower than about $30$~GeV.

Finally, taking into account the effects discussed so far, we find the baryon-to-entropy ratio to be
\bea
\frac{n_B}{s} \simeq 
\frac{1}{\Delta} 
\frac{2.6\,\epsilon\times 10^{-10} }{r\sin\alpha}  
\left(\frac{T_n}{60\,{\rm GeV}}\right)^2,
\eea
where we have taken $g_\ast=100$, and $\Delta\simeq (T_{\rm reh}/T_n)^3\,e^{K_\phi}$ 
describes the dilution of baryon number. 
The above relation shows that spontaneous EWBG induced by the ALP can account for the observed 
baryon asymmetry, $n_B/s\simeq 8.6\times 10^{-11}$, if $\Delta$ lies in the range
\bea
1 \leq \Delta \lesssim 10,
\eea
which is the case for $T_n$ above about $30$~GeV and $f$ below $10^8$~GeV.
The dilution factor exponentially increases for larger $f$ or in the region with $v_n/T_n<1$.
It should be noted that $T_n$ is close to $T_2$, and the dependence of $f$ of the baryon 
asymmetry comes in through the washout factor.
Thus, in the small washout regime with $\Delta$ below about 10, the baryon asymmetry 
becomes not very dependent on $f$.

Let us show the evolutions of relevant physical quantities on figures.
In the left panel of Fig.~\ref{fig:evolution}, the curves show how the Higgs background field value 
evolves for $f$ between $10^6$ and $10^7$~GeV in the case with 
\bea
\alpha = 1.4,\quad \epsilon=0.95,\quad r=1.1,
\eea
for which $T_2\simeq 84$~GeV. 
The scalar potential is asymmetric about the minimum $v=v_n$ for nonzero
$\alpha$ and $\epsilon$, and sphalerons are active only in the region below the lower horizontal 
dashed line, where $v<v_{\rm cut}$. 
One can see that the number of relevant oscillations decreases with $f$. 
The right panel shows the evolution of the baryon number.
The baryon number is produced at the first falling, and then is washed out if later 
oscillations pass the region of rapid sphaleron transitions.

\begin{figure}[t] 
\begin{center}
\includegraphics[width=0.42\textwidth]{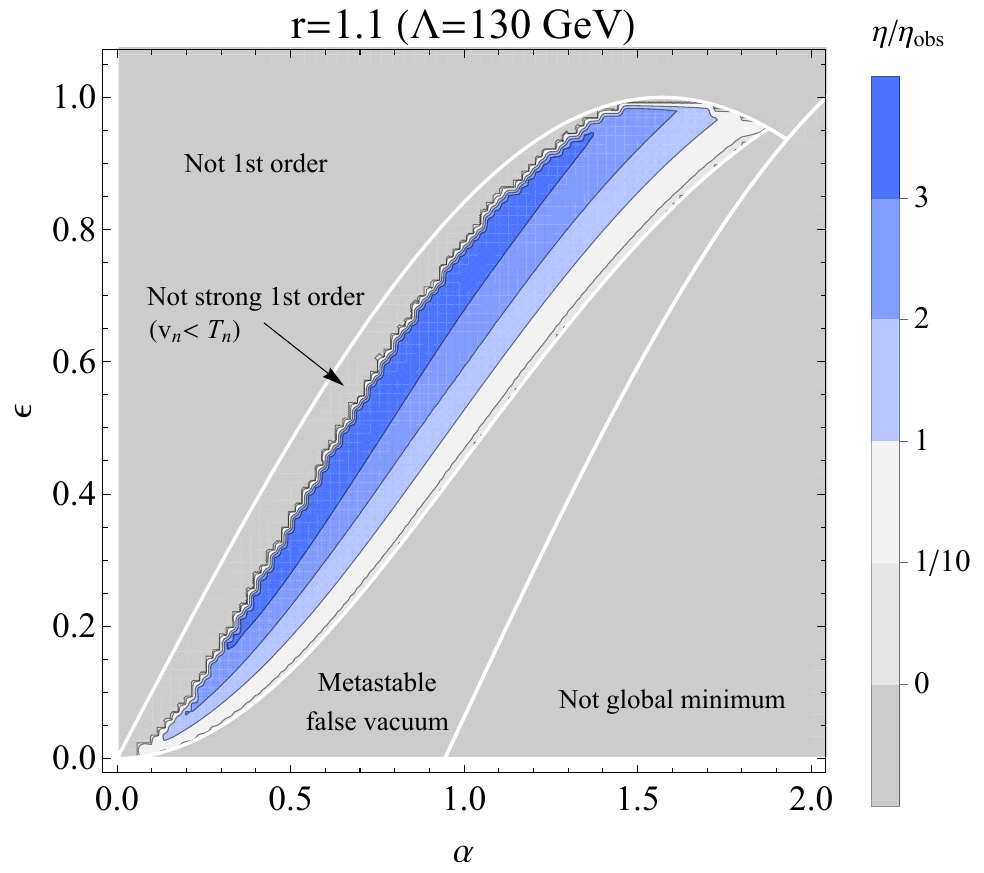} 
\end{center}
\caption{
Spontaneous EWBG realized by the ALP in the case with $r=1.1$ and $f=10^6$~GeV. 
The gradient represents the relic baryon number density normalized by the observed value. 
The correct baryon asymmetry is obtained in the blue shaded region for an appropriate
dilution factor.
Sizable dilution requires $f$ above $10^7$~GeV, below which the baryon asymmetry 
becomes insensitive to $f$.
 }
\label{fig:final}
\end{figure}

We close this section by examining the viable region of parameter space for the case with
\bea
r=1.1,  \quad f=10^6\,{\rm GeV}.
\eea
The blue shaded region in Fig.~\ref{fig:final} leads to the correct baryon asymmetry,
and the color gradient represents the required value of dilution factor $\Delta$.
Note that sizable dilution is obtained for $f$ above $10^7$~GeV. 
Here we have used that $T_n$ is close to $T_2$, which does not depend on $f$,
and that the region for a first-order phase transition with $v_n/T_n>1$ does not change much
with $f$ because $V_{\rm eff}(\theta)$ relies on $f$ only via small radiative corrections. 
This also indicates that the viable region would remain almost the same 
if one considers smaller $f$.

\section{Experimental constraints}
\label{sec:EC}

In this section, we summarize experimental constraints on the ALP. 
The coupling to the Higgs mass squared operator makes the ALP mix with the Higgs boson,
and thus there are various constraints depending on the mixing angle
\bea
\sin\delta \simeq
 \frac{r^2 \sin\theta_0}{2}\times\frac{v_0}{f}, 
\label{mixing}
\eea
and its mass
\bea
m_\phi \simeq
\sqrt{\frac{r^2(\sin\alpha +   r^3\epsilon  \sin^3\theta_0)}{4\sin(\theta_0+\alpha)}}
\times\frac{v_0m_h}{f}, 
\eea
where $m_h\simeq 125$~GeV is the Higgs boson mass.
First, the ALP is subject to the bound on the EDM 
because its coupling to the EW anomaly violates $CP$ symmetry in the presence of mixing with the Higgs boson.
%
%{\color{magenta}  because combined with Higgs mixing, its coupling to EW anomaly  violates CP symmetry.} 
The electron EDM is radiatively generated as~\cite{Choi:2016luu}
\bea
d_e  &\simeq& 
\frac{8e^3}{(4\pi)^4}
\frac{m_e}{v_0}\frac{\sin\delta}{f}\ln\left(
\frac{m_h}{m_\phi}\right) 
\nonumber \\
&\sim& 10^{-34}\, e\,{\rm cm} \times \left(\frac{10^6\,{\rm GeV}}{f}\right)^2, 
\label{edm}
\eea
where $m_e$ is the electron mass. 
If $f$ is larger than about $5$~TeV, the above contribution is below the latest experimental 
bound from ACME II in the region of parameter space for a strong first-order phase transition. 
One may naively expect that such a large $f$ would also suppress $CP$-violating effects 
on baryogenesis since the ALP is responsible for the CS chemical potential. 
However, as shown in Sec.~\ref{sec:BG}, the baryon asymmetry is generated depending on
how rapidly $\phi/f$ changes during EWPT.
The ALP excursion $\Delta \phi$ is of the order of $\alpha f$ during EWPT, and thus 
spontaneous EWBG can work at $f$ much above TeV while being free from the EDM constraints.
On the other hand, other EWBG scenarios
%{\color{magenta} Since the ALP is responsible for the CS chemical potential, a naive expectation is that the CPV for EWBG is also suppressed as $f$ increases. However as we show in Sec.~\ref{sec:BG}, the CPV for EWBG is proportional to 
%the change of $\phi/f$ during EWPT and $\Delta \phi = {\cal O}(f)$ provides much enhanced effect compared to the naive expectation. 
%Spontaneous EWBG working at $f$ much above TeV is totally free from
%the EDM constraints.  
%On one hand,}
%other scenarios for EWBG 
generally suffer from the EDM constraints because the Higgs sector
is modified by a singlet scalar significantly coupled to it to induce a strong first-order phase
transition. 
For instance, another simple candidate for a time-dependent EW theta would be
$\Theta_{\rm EW} = |H|^2/\Lambda^2_{\rm cut}$, where $\Lambda_{\rm cut}$ is the cutoff scale 
of the effective coupling.  
Then, baryon asymmetry is produced during phase transition according to 
$n_B \propto v^2_{\rm cut}/\Lambda^2_{\rm cut}$ in the adiabatic limit. 
The correct baryon asymmetry requires $\Lambda_{\rm cut}$ lower than $0.5$~TeV 
if the phase transition occurs around the EW scale. 
However, the latest bound on electron EDM from ACME II excludes $\Lambda_{\rm cut}$ 
below about $6\times 10^5$~GeV.

Our scenario solves the matter-antimatter asymmetry problem while avoiding 
the electron EDM bound in the weakly coupled regime with $f$ between about $5$~TeV 
and $10^8$~GeV.
This corresponds to the ALP mass in the range between sub MeV and $5$~GeV, 
for which stringent constraints come from rare meson decays and also from beam-dump 
ALP searches~\cite{Flacke:2016szy}. 
In addition, if lighter than about $20$~MeV, which is roughly the supernova temperature,  
ALPs can be produced in supernovae.
Supernova cooling is accelerated if the produced ALPs efficiently escape from it,
implying that the ALP-Higgs mixing should lie in a certain range to avoid conflict with
the observation.

Let us describe the experimental constraints from meson decays in more detail.
If the ALP has a mass in the range between $2m_\mu$ and $m_B-m_K$, 
where $m_i$ denotes the mass of the indicated particle, 
the mixing should be suppressed to be consistent with the limit on the decay rate 
for $B \to K \phi \to K \mu^+\mu^-$ obtained at Belle and LHCb~\cite{Aaij:2012vr,Wei:2009zv,Aaij:2015tna, Bezrukov:2014nza, Hyun:2010an}.
The electron channel, $B\to K\phi \to K e^+e^-$, gives a weaker constraint than the muonic one. 
On the other hand, in the case with $m_\phi < m_K - m_\phi$, the mixing is constrained
mainly by rare $K$ meson decays.
Especially, if $m_\phi< 2m_\mu$, the ALP is subject to a stringent bound imposed by 
the searches for invisible $K$ meson decays at BNL E787 and E949 experiments~\cite{Adler:2004hp,Artamonov:2009sz}.

 \begin{figure}[t]
   \includegraphics[width=0.38\textwidth]{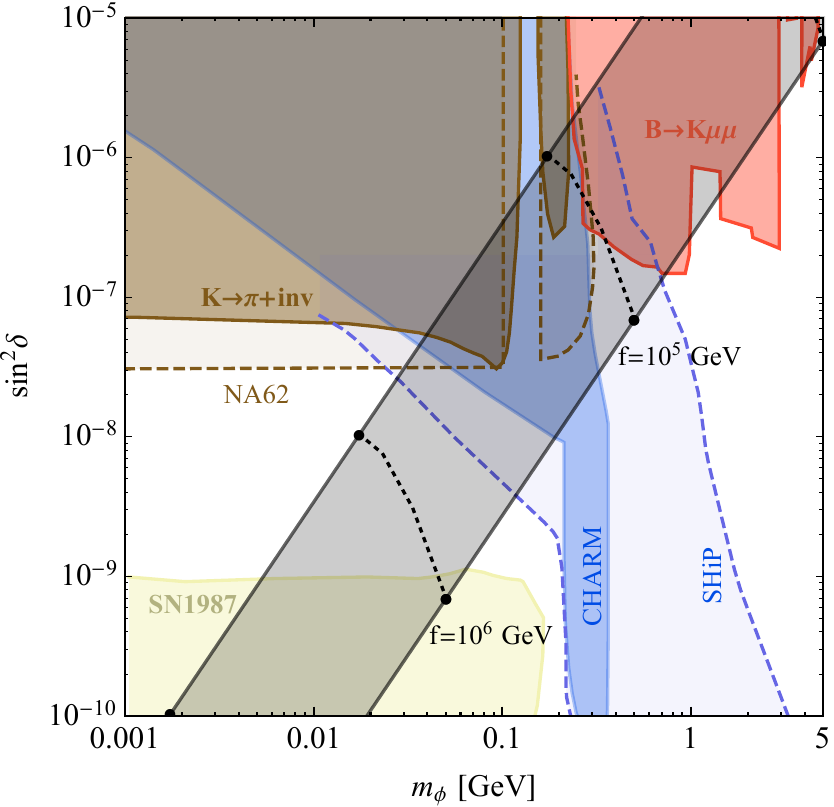} 
   \caption{
   Experimental constraints on the ALP properties from rare meson decays, supernova cooling,
   and beam-dump ALP searches~\cite{Flacke:2016szy}. 
   Here $\delta$ is the ALP-Higgs mixing angle, and the ALP is assumed not to decay 
   into hidden particles. 
   The light blue and purple regions will be reached in future experiments.
   The gray band shows the relation between the ALP mass and mixing for $r=1.1$ 
   by taking $0.1\leq \alpha \leq 1.4$ and $0.14\leq \epsilon \leq 0.8$.
   The black dotted lines on the band are the contours of $f$.  
   }
   \label{fig:pheno}
 \end{figure}

In Fig.~\ref{fig:pheno}, we summarize the current experimental constraints on the ALP properties 
for $1\,{\rm MeV}\lesssim m_\phi \lesssim 5\,{\rm GeV}$.
The dark purple and cyan regions are excluded by rare $K$ and $B$ meson decays, respectively.
The sky blue region leads to too rapid supernova cooling, while the red region is excluded 
by the beam-dump ALP searches at CHARM~\cite{Bergsma:1985qz}.
One can see that the viable window is
\bea 
10^{-9} \lesssim \sin^2\delta \lesssim 3 \times 10^{-7},
\eea
for $1\,{\rm MeV}\lesssim m_\phi \lesssim 0.2\,{\rm GeV}$,
while it is 
\bea 
{\rm Br}(\phi \to \mu^+\mu^-)\times \sin^2\delta \lesssim 
6\times 10^{-7},
\eea
for $0.3\,{\rm GeV}\lesssim m_\phi \lesssim 5\,{\rm GeV}$, 
where ${\rm Br}(\phi\to \mu^+\mu^-)$ is the branching ratio for the ALP decay into 
a muon pair. 
Here we have assumed that the ALP does not decay into hidden sector particles.
If allowed, the constraint from $B$ meson decays will be weakened.
In our scenario, the approximate relation 
\bea
m_\phi \sim m_h \sin\delta,
\eea holds
between the ALP mass and mixing angle. 
The gray band shows such relation for $r=1.1$ in the parameter space, 
$0.1\leq \alpha \leq 1.4$ and $0.14\leq \epsilon \leq 0.8$.
Thus, a viable region appears for $m_\phi$ in the MeV to GeV scale,
or equivalently $f$ in the range between about $10^5$ and $10^7$~GeV.
We also plot the constant contours of $f$ on the band by black dotted lines.
It is interesting to note that the light blue and brown regions will be probed by
experiments at SHiP~\cite{Alekhin:2015byh} and NA62, respectively.

\section{Conclusions}
\label{sec:Con} 

In this paper, we have shown that an ALP provides a simple and natural framework for EWBG
in a wide range of $f$ owing to its periodic nature. 
In particular, for $f$ much above TeV, it offers a new direction in which the EDM and the LHC are 
no longer a probe of EWBG.   
Instead, ALP searches would reveal the interesting connection between EW symmetry breaking 
and baryogenesis established in our scenario. 
A part of the parameter space is already constrained by the existing results 
from ALP searches.

The ALP triggers a strong first-order phase transition insensitively to the value of $f$, and leads to
the adiabatic regime at $f$ above $10$~TeV, where thick bubble walls prevent non-local
baryon production from particle diffusion. 
Nonetheless, coupled to the EW anomaly, the ALP can naturally realize local spontaneous EWBG to 
solve the matter-antimatter asymmetry problem
for $f$ below $10^8$~GeV while avoiding strong washout. 
Interestingly, the phase transition occurs smoothly because the bubble nucleation temperature
is close to the bubble disappearing temperature, 
and baryogenesis proceeds almost isothermally.

Using that the ALP-Higgs mixing is constrained by various experiments, we find 
the viable window to be $f$ from about $10^5$ to $10^7$~GeV, which corresponds 
to ALP mass in the MeV to GeV scale. 
In such a weakly coupled regime, our scenario is completely safe not only from the EDM 
constraints, especially from the bound on the electron EDM recently improved by ACME II,  
but also from Higgs precision measurements. 
These are the features distinguishable from the conventional scenarios of EWBG.
The indicated ALP window, which has suffered from the lack of strong theoretical interest, could be probed 
in future beam-dump experiments such as SHiP.
\\
\\

\noindent{\bf Acknowledgments}
THJ thanks G\' eraldine Servant and Eibun Senaha for useful discussions.
This work was supported by IBS under the project code, IBS-R018-D1 (THJ and CSS), 
and by the National Research Foundation of Korea (NRF) grant funded by the Korea
government (MSIP) (NRF-2018R1C1B6006061) (KSJ).  T. H. J. is also supported 
by the US Department of Energy grant DE- SC0010102 and 
Prof. Kohsaku Tobioka's startup fund at Florida State University (Project id: 084011- 550- 042584).

\appendix\section*{Appendix}

\setcounter{section}{1}
\setcounter{equation}{0}

\subsection{Bounce solution} 
\label{appendix:bounce}

In this appendix we analyze the structure of the bounce solution in more detail
and discuss how the tunneling proceeds.
The Higgs and ALP fields are fixed at the symmetric false vacuum,
$(\phi,h)=(0,0)$, at high temperatures. 
As temperature drops, the potential develops a new minimum at
$(\phi,h)\neq (0,0)$,
and the false vacuum can decay via nucleation of critical bubbles of the broken phase
if it has a higher free energy than the EW vacuum.  
The nucleation rate per unit volume is
\bea 
\Gamma  \propto T^4 e^{ - S_3/T}, 
\eea 
where the Euclidean action for the bounce is given by  
\bea 
\hskip -0.2cm S_3 = 4\pi \int_0^\infty dr\, r^2 \left[ 
\frac{1}{2}
\left(\frac{d h}{d r}\right)^2
  + \frac{1}{2}\left(\frac{d\phi}{dr}\right)^2  
+ V  \right],
\eea 
for the scalar potential $V$ at a temperature $T$.
Here we have set $V=0$ at the symmetric vacuum.
The bounce field configuration can be found from
\bea
\frac{d^2 \varphi}{dr^2} + \frac{2}{r}\frac{d \varphi}{d r} = 
\partial_\varphi V,  
\eea
under the boundary conditions,
$d\varphi/dr=0$ at $r=0$ and $\varphi=0$ at $r=\infty$,  
where $\varphi=\{\phi,h\}$.

It is usually the case that one needs to take numerical calculations to find the bounce
solution.  
However, in our scenario, a large $f$ allows an analytic approach because bubble
nucleation occurs mostly along the light field direction and the potential is insensitive
to $f$ for a given $\theta$, where $\theta=\phi/f$. 
This feature becomes transparent when the action is written in the form
\bea
\label{eq:S3_2}
S_3 = 4\pi f^3 
\int ^\infty_0
dx\, x^2
\left[ \frac{h'^2}{2 f^2} + \frac{\theta'^2}{2} + V(h,\theta)\right], 
\eea
where $x\equiv r/f$, and the prime is the derivative with respect to $x$.
The equations of motion then read
\bea
\label{eq:eom_bounce}  
\frac{1}{f^2}\left(
h^{\prime\prime} + \frac{2}{x} h^\prime \right)
&=& \partial_h V,  
\nonumber \\ 
\theta^{\prime\prime} + \frac{2}{x} \theta^\prime
&=& \partial_\theta V.  
\eea
For the potential (\ref{potential}), field variations over the critical bubble
are roughly given by $\Delta h \sim \Lambda$ and $\Delta \theta \sim \alpha$. 
Using the equation of motion for $\theta$, one can estimate the size of the bubble 
to be $\Delta x\sim 1/\Lambda^2$ because the insensitivity of $V(h,\theta)$ to $f$ implies
\bea
\Delta \theta\,\partial_\theta V \sim \alpha \Lambda^4.
\eea 
Combined with these relations, the equation of motion for $h$ leads to 
\bea
\Delta h\, \partial_h V \sim \left(\frac{\Lambda}{f}\right)^2
\Lambda^4.
\eea 
Therefore, for $\Lambda\ll f$, the Higgs trajectory for the bounce is effectively fixed 
by $\partial_h V=0$, making the ALP feel a potential along it.  
This justifies why the tunneling can be examined within the effective theory of
the light ALP constructed by integrating out the heavy field $h$:
\bea
S_3 \simeq  4\pi f^3 \int^\infty_0 d x\, x^2\left[
\frac{\theta'^2}{2} + V_{\rm eff}(\theta) \right],
\eea
where 
$V_{\rm eff}(\theta)  = V(\hat h(\theta),\theta)$ for the Higgs field value
$h=\hat h(\theta)$ satisfying $\partial_h V=0$. 
The bounce solution is obtained from
\bea
\theta^{\prime\prime} + \frac{2}{x} \theta^\prime = \partial_\theta V_{\rm eff},
\eea
under the boundary conditions, $\theta^\prime=0$ at $x=0$ and $\theta=0$ at $x=\infty$.
Note that the equation of motion is independent of $f$, implying $S_3 \propto f^3$
for a given temperature.

Bubble nucleation happens within a Hubble time if $S_3/T \lesssim 140$. 
In our scheme, because of the large prefactor $f^3$ in $S_3$, the nucleation starts when 
the universe cools down close to $T=T_2$ so that the barrier of the potential is low enough.
Here $T_2$ is the temperature at which the barrier between minima of $V_{\rm eff}$ disappears. 
At a temperature near $T_2$, the effective potential around $\theta=0$ can be 
approximated as 
\bea
\hspace{-0.6cm}
\frac{2 V_{\rm eff}}{\Lambda^4}
=
\left\{
\begin{array}{ll}
\theta^2 + {\cal O}(\theta^4) & {\rm for}\,\,\theta>-\theta_\ast \\
\theta^2  -\kappa^2(\theta + \theta_\ast)^2 + {\cal O}(\theta^3)  & 
{\rm for}\,\,\theta<-\theta_\ast \\
\end{array}
\right.
\eea 
for $\kappa$ and $\theta_\ast$ depending on $T$ and the model parameters.
Here, $\kappa$ is larger than unity, and $\theta_\ast$ is small and proportional
to $(T/T_2 - 1)\sin\alpha$. 
It then follows that the curvature of the potential changes sign at $\theta=-\theta_\ast$.  
For $T>T_2$, one also finds that $V_{\rm eff}=0$ at $\theta=0$ and 
$\theta=-\kappa\, \theta_\ast/(\kappa-1)$, and there is a potential barrier lying 
between the two points.

Let us now examine the bounce solution, which relies on the potential shape between the
two points giving $V_{\rm eff}=0$ at $T>T_2$.
The equation of motion can be solved analytically because $\partial_h V_{\rm eff}$ is 
approximately linear in $\theta$ in the relevant region. 
Introducing a dimensionless variable for simplicity
\bea 
\label{eq:variables}
\rho \equiv  \sqrt{\kappa^2 - 1} \Lambda^2 x,  
\eea
we find that the solution is written
\bea
\tilde\theta 
\equiv \frac{\theta}{\theta_\ast}
\simeq 
\left\{
\begin{array}{ll}  
 \frac{\kappa^2}{\kappa^2-1} 
+ c_1 \frac{\sin \rho}{\rho}   & {\rm for}\   \rho < \rho_0 \\ 
1 & {\rm for} \ \rho = \rho_0 \\
   c_2 \frac{ \exp(- \rho/\sqrt{\kappa^2-1})}{\rho}  &  {\rm for}\   \rho > \rho_0  
\end{array}
\right., 
\eea 
where the coefficients $c_1$ and $c_2$ are given by
\bea 
c_1 &=& \frac{\kappa^2 (\rho_0 + \sqrt{\kappa^2 - 1})}{(\kappa^2 - 1) 
(\sqrt{\kappa^2 - 1} \cos \rho_0 + \sin \rho_0)}, 
\nonumber \\
c_2 &=&\frac{ \kappa^2 \exp(\frac{\rho_0}{\sqrt{\kappa^2-1}}) (\rho_0 \cos \rho_0 - \sin \rho_0)}{(\kappa^2 -1)\cos \rho_0 
+\sqrt{\kappa^2 - 1} \sin \rho_0},
\eea 
which follows from the fact that $\theta$ and its derivative are continuous at $\rho=\rho_0$
with $\rho_0$ fixed by the condition
\bea 
 \frac{\kappa^2 (\rho_0 \cos \rho_0 - \sin \rho_0)}{ (\kappa^2 -1)\rho_0\cos \rho_0 + \sqrt{\kappa^2 - 1} \rho_0\sin \rho_0} = 1.
\eea  
Note that $\rho_0$ is about $1.43\pi$ in the limit $\kappa\to1$, and it monotonically decreases 
with $\kappa$ while approaching to $\pi$. 
For instance, $\rho_0\simeq 1.18\pi$ at $\kappa=\sqrt2$.

\begin{figure}[t]
 \begin{center}
 	\includegraphics[width=0.37\textwidth]{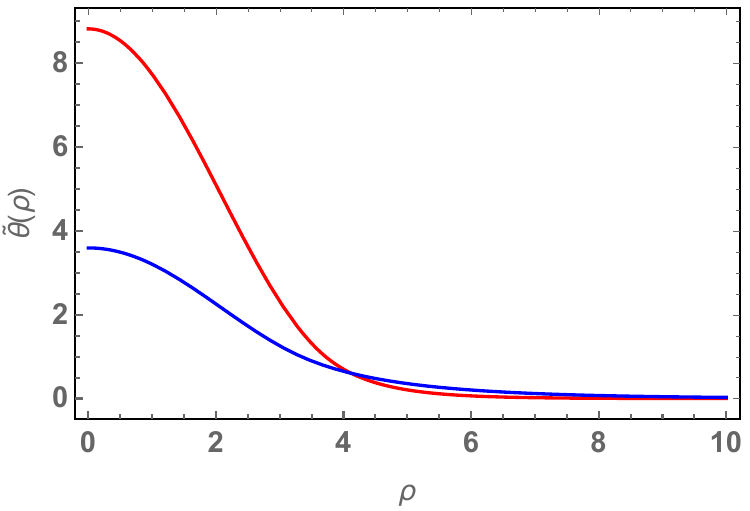} 
 \end{center}
 \caption{
Bounce solution for $\kappa=\sqrt{2}$ (red) and $2\sqrt{2}$ (blue).
 }
\label{fig:bounce}
\end{figure}

Fig.~\ref{fig:bounce} illustrates the profile of $\tilde\theta$ as a function of $\rho$. 
As one can see in the figure, the critical bubble has a thick wall because the field varies smoothly
within the bubble of radius
\bea
R_c = \frac{f}{\sqrt{\kappa^2-1} \Lambda^2} \rho_0
\sim 
\frac{\pi}{\sqrt{\kappa^2-1}} \frac{f}{\Lambda^2}.
\eea
The radius of the critical bubble in the real space can be regarded as
the bubble wall width $L_w$ in the rest frame of the expanding bubble 
wall,~i.e.~$L_w\sim R_c$.

Finally, using the results obtained so far, one can estimate 
the bounce exponent around $T=T_2$ to be
\bea
\frac{S_3}{T} \simeq
c(\kappa)
\frac{4\pi f^3 \theta^2_\ast}{T_2 \Lambda^2},
\eea
where $c(\kappa)$ is numerically calculable for a given $\kappa$, 
which is typically of order unity.
For instance, $c(\sqrt2)\simeq 23$, $c(2\sqrt2)\simeq 1.5$ and 
$c(7)\simeq 0.33$. 
It is important to notify that $S_3/T$ is proportional to $(T-T_2)^2f^3$ because 
$\theta_\ast \propto (T/T_2 -1)$. 
Thus, in Eq.~(\ref{eq:S3}), $n$ is equal to $2$.

\subsection{Residual field oscillations inside a bubble} \label{appendix:res_osc}

The EW vacuum $(\phi,h) = (\phi_T,v_T)$ at $T$, which is determined  by 
$\partial_\phi V = \partial_h V = 0$, has 
\bea
-\alpha f \lesssim \phi_T <0, \quad
0< v_T \lesssim \Lambda,
\eea 
and becomes the true vacuum when $T$ drops down below the critical temperature.
Just after a critical bubble is formed at the nucleation temperature $T=T_n$, 
the Higgs and ALP fields take values 
\bea
\phi(t_n) \sim  \theta_\ast(t_n) f  \quad{\rm and}\quad
h(t_n) \ll v_T,
\eea  
inside the bubble, and they are initially located far from the true vacuum.
Here we have used the fact that $\theta_\ast(t_n)$ is much smaller than $\alpha$ in size because
$\theta_\ast$ is proportional to $(T/T_2-1)\sin\alpha$ and $T_n$ is close to $T_2$.
Therefore, as the bubble expands, the fields classically roll toward the true vacuum while oscillating
about it. 
Their evolution can be understood by looking deep inside the bubble, where the effect of spatial gradients
are small. 
For a time scale much shorter than the Hubble time, the field evolution can be approximated by 
\bea 
&&
\ddot{h} + \Upsilon_h  \dot{h} + \partial_h V = 0,
\nonumber \\
&&
\ddot{\phi} +  \Upsilon_\phi \dot\phi+  \partial_\phi V = 0,  
\eea 
%where the dot denotes a derivative with respect to time $t$.  
where 
 the thermal
dissipation rate $\Upsilon_h$ is determined by  Higgs interactions to the background
thermal plasma especially from top quark contribution~\cite{Mukaida:2012qn}, 
\bea 
\Upsilon_h \sim 
\left\{
  \begin{array}{ll r}
    y^2_t \Gamma_{\rm th}     &  {\rm for} &    y_t h\lesssim \Gamma_{\rm th} \\
    y^2_t \frac{(y_t h)^2}{\Gamma_{\rm th}}    &  {\rm for}   &  
    y_t h \gtrsim \Gamma_{\rm th}
  \end{array}
  \right.,
\eea
which includes an uncertainty of order unity due to the complicated dispersion relation 
for the top quark~\cite{Wang:1999mb}.
Note also that the weak gauge bosons give important contributions when the Higgs background
field value $h$ is sizable. 
%
%{\color{red} 
%\bea
%\Upsilon_h &\sim &
% y_t^2 \alpha_{t, g} T  \quad\ \, {\rm for} \quad
% y_t h \lesssim \alpha_{t,g} T \nonumber\\
% &\sim & y_t^2 \frac{(y_t h)^2}{\alpha_{t, g} T}   
% \quad {\rm for} \quad y_t h \gtrsim \alpha_{t, g} T 
%\eea 
% There are  ${\cal O}(1)$ uncertainties for the estimation, because 
% the rather complicated dispersion relation for top quarks \cite{Wang:1999mb}. 
%Contributions from the weak gauge bosons become important  when $h$ is sizable.}
On the other hand, the anomalous coupling in Eq.~(\ref{anomaly}) acts as a friction term,
dissipating the ALP energy non-perturbatively  with~\cite{McLerran:1990de,Burnier:2005hp,DOnofrio:2014rug}
\bea
\Upsilon_\phi 
=
\frac{\Gamma_{\rm sph}}{T f^2}  \sim 10^{-6} \frac{T^3}{f^2},
\eea
for $T > E_{\rm sph}$, and 
\bea
\Upsilon_\phi 
\sim 10^{-5}
\frac{T^3}{f^2} \left(\frac{2m_W}{\alpha_W T}\right)^7 
e^{- \frac{E_{\rm sph}}{T}},
\eea 
for $T < E_{\rm sph}$, where the mass of the weak gauge boson and the sphaleron energy 
depend on the Higgs background field value as
$m_W= g h/2$ and $E_{\rm sph}\simeq 4\pi h/g$, respectively. 
The trilinear ALP-Higgs-Higgs coupling can also give a sizable contribution to
$\Upsilon_\phi$ during the period when the Higgs boson mass is comparable to or smaller than
the temperature.  
%also could be important, because 
%the Higgs mass is the same order of the temperature.

Subject to large thermal friction, the Higgs field is expected to be quickly frozen to the vacuum value 
$v_T$ within a time scale $1/\Upsilon_h \sim 1/T$. 
However, there is a residual oscillation induced by its mixing with the ALP,
which is sizable even for a tiny mixing because the ALP has a large field excursion during 
its evolution.  
To examine qualitatively the field evolution, we take a quadratic approximation of the potential
around the true vacuum
\bea
V \simeq 
\frac{m^2_L}{2}\varphi^2_L + \frac{m^2_H}{2} \varphi^2_H,
\eea
for the light and heavy mass eigenstates given by 
%where $\varphi_L$ and $\varphi_H$ are the 
%{\color{red}  light and heavy} mass eigenstates, respectively, given by 
\bea
\varphi_L &=& \Delta \phi \cos\delta + \Delta h \sin\delta, 
\nonumber \\
\varphi_R &=& \Delta h \cos\delta - \Delta \phi \sin\delta, 
\eea 
which respectively have masses
$m_L\simeq m_\phi \sim \Lambda^2/f$ and $m_H \simeq m_h \sim \Lambda$ at a given 
temperature.
Here  $\Delta h \equiv h - v_T$ and $\Delta \phi \equiv \phi - \phi_T$ are the displacements
from the true vacuum, and the mixing angle is roughly given by  $\sin\delta\sim m_\phi/m_h$. 
It is straightforward to obtain the equations of motion in the canonical basis
\bea
&&
\hspace{-0.7cm}
\left(\frac{d^2}{dt^2}
+ \Upsilon_h \sin^2\delta\frac{d}{dt} 
+ m^2_L
\right) \varphi_L \simeq  -\frac{\sin2\delta}{2}\Upsilon_h \dot\varphi_H,
\nonumber \\
&&
\hspace{-0.7cm}
\left(\frac{d^2}{dt^2}
+ \Upsilon_h \cos^2\delta\frac{d}{dt} 
+ m^2_H
\right) \varphi_H \simeq -\frac{\sin2\delta}{2}\Upsilon_h \dot\varphi_L,
\eea 
with the initial conditions
\bea
\varphi_L(t_i) &\simeq& -\phi_T\cos\delta - v_T\sin\delta \sim f,
\nonumber \\
\varphi_H(t_i) &\simeq& -v_T\cos\delta +  \phi_T\sin\delta \sim -\Lambda.
\eea 
Here we have ignored the effect of the anomalous coupling on the motion of $\varphi_L$
because $\Upsilon_\phi$, which is smaller than about $10^{-6} T^3/f^2$, is  
much suppressed compared to the mixing-induced friction term $\Upsilon_h \sin^2\delta \sim T v^2/f^2$
in the whole range of $h$ we are interested in.  
%{\color{blue} 
%Here we ignore the effect of dissipation by sphaleron process ($\Upsilon_\phi$) on the motion of $\varphi_L$. The reason is that 
%the effect of $\Upsilon_\phi\lesssim 10^{-6} T^3/f^2$ should be compared with that of $\Upsilon_h\sin^2\delta$,  and $\Upsilon_h\sin^2\delta \sim  T m_\phi^2/m_h^2\sim T v^2/f^2 \gg \Upsilon_\phi$ for the whole range of $h$ we are interested in. } 
In the equation of motion for the heavy field $\varphi_H$,  
the term proportional to $\dot\varphi_L$ is much smaller than
$\Lambda^3$ in size and so can be neglected until $\varphi_H$ 
gets close to its vacuum value.
The thermal friction term thus quickly freezes
$\varphi_H$ to the vacuum value within a time scale $1/T$.
For a time scale much shorter than the Hubble time, the solutions are approximated to be
\bea
\varphi_L &\approx&
\phi_T\, e^{-\delta^2 \Upsilon_h t/2}
\cos\Big(
\sqrt{m^2_\phi-\delta^4\Upsilon^2_h}\,t + \beta_L\Big),
\nonumber \\
\varphi_H &\approx&
v_T\, e^{-\Upsilon_h t/2}
\cos\Big(
\sqrt{m^2_h - \Upsilon^2_h}\,t + \beta_H \Big),
\eea
at temperatures below $T_n$, where $\beta_L$ and $\beta_H$ are constant phases.
We have numerically confirmed the above approximations. 
Note also that the Higgs field evolves according to 
\bea
\hspace{-0.6cm}
h(t) \approx
v_T
- \phi_T \sin\delta \times e^{-\delta^2 \Upsilon_h t/2}
\cos(m_\phi t + \beta_L),
\eea
which follows from $\Delta h\simeq \varphi_L \sin\delta$ for $t\gg 1/T$. 
Because $\phi_T \sin\delta \sim \Lambda$, the residual oscillation of the Higgs field can be 
sizable for a time scale less than about $1/(\delta^2 T)$.
Fig.~\ref{fig:evolution}, which is obtained via a numerical calculation, illustrates such a feature.

The results above can be understood more easily by replacing the heavy Higgs field $h$ 
with $\hat h(\phi)$ because the rapid damping $\varphi_H\to 0$
means $\partial V/\partial h \to 0$ for a given value of $\phi$. 
In such constructed effective theory, the ALP oscillates about the true minimum $\phi=\phi_T$ after tunneling, 
and accordingly the Higgs background field value changes because it is given by $v(t)=\hat h(\phi(t))$.
Note also that  the ALP effectively couples to SM particles with a coupling proportional to 
$\partial \hat h/\partial\phi|_{\phi_T}$, and thus its oscillating energy is thermally
dissipated.

For a final remark, we note that the real situation is more complicated because the mixing
angle is field-dependent, and self interactions during bubble expansion are also important. 
Our point here is that there can be sizable Higgs oscillations, which would then wash out 
the baryon asymmetry.
In the analysis, we have taken into account such effects to obtain a conservative estimation
of the final baryon abundance.


\begin{thebibliography}{99}
 %\cite{Andreev:2018ayy}
\bibitem{Andreev:2018ayy} 
  V.~Andreev {\it et al.} [ACME Collaboration],
  %``Improved limit on the electric dipole moment of the electron,''
  Nature {\bf 562}, no. 7727, 355 (2018).
 % doi:10.1038/s41586-018-0599-8
  %%CITATION = doi:10.1038/s41586-018-0599-8;%%
  %9 citations counted in INSPIRE as of 06 Nov 2018
 
 
 \bibitem{Bian:2014zka} 
  L.~Bian, T.~Liu and J.~Shu,
  %``Cancellations Between Two-Loop Contributions to the Electron Electric Dipole Moment with a CP-Violating Higgs Sector,''
  Phys.\ Rev.\ Lett.\  {\bf 115}, 021801 (2015)
  %doi:10.1103/PhysRevLett.115.021801
  [arXiv:1411.6695 [hep-ph]].
  %%CITATION = doi:10.1103/PhysRevLett.115.021801;%%
  %31 citations counted in INSPIRE as of 05 Sep 2019 

 
 
 \bibitem{Cline:2017qpe} 
  J.~M.~Cline, K.~Kainulainen and D.~Tucker-Smith,
  %``Electroweak baryogenesis from a dark sector,''
  Phys.\ Rev.\ D {\bf 95}, no. 11, 115006 (2017)
 % doi:10.1103/PhysRevD.95.115006
  [arXiv:1702.08909 [hep-ph]].
  %%CITATION = doi:10.1103/PhysRevD.95.115006;%%
  %17 citations counted in INSPIRE as of 05 Sep 2019


%\cite{Baldes:2018nel}
\bibitem{Baldes:2018nel} 
  I.~Baldes and G.~Servant,
  %``High scale electroweak phase transition: baryogenesis & symmetry non-restoration,''
  JHEP {\bf 1810}, 053 (2018)
  %doi:10.1007/JHEP10(2018)053
  [arXiv:1807.08770 [hep-ph]].
  %%CITATION = doi:10.1007/JHEP10(2018)053;%%
  %7 citations counted in INSPIRE as of 08 Sep 2019

 
 
%\cite{Jeong:2018ucz}
\bibitem{Jeong:2018ucz} 
  K.~S.~Jeong, T.~H.~Jung and C.~S.~Shin,
  %``Axionic Electroweak Baryogenesis,''
  Phys.\ Lett.\ B {\bf 790}, 326 (2019)
  %doi:10.1016/j.physletb.2019.01.036
  [arXiv:1806.02591 [hep-ph]]. 
 
 
 
 
 %\cite{Jaeckel:2015jla}
\bibitem{Jaeckel:2015jla} 
  J.~Jaeckel and M.~Spannowsky,
  %``Probing MeV to 90 GeV axion-like particles with LEP and LHC,''
  Phys.\ Lett.\ B {\bf 753}, 482 (2016)
  %doi:10.1016/j.physletb.2015.12.037
  [arXiv:1509.00476 [hep-ph]].
  %%CITATION = doi:10.1016/j.physletb.2015.12.037;%%
  %56 citations counted in INSPIRE as of 07 Nov 2018
  
 %\cite{Knapen:2016moh}
\bibitem{Knapen:2016moh} 
  S.~Knapen, T.~Lin, H.~K.~Lou and T.~Melia,
  %``Searching for Axionlike Particles with Ultraperipheral Heavy-Ion Collisions,''
  Phys.\ Rev.\ Lett.\  {\bf 118}, no. 17, 171801 (2017)
  %doi:10.1103/PhysRevLett.118.171801
  [arXiv:1607.06083 [hep-ph]].
  %%CITATION = doi:10.1103/PhysRevLett.118.171801;%%
  %31 citations counted in INSPIRE as of 07 Nov 2018 
 
  
%\cite{Mariotti:2017vtv}
\bibitem{Mariotti:2017vtv} 
  A.~Mariotti, D.~Redigolo, F.~Sala and K.~Tobioka,
  %``New LHC bound on low-mass diphoton resonances,''
  Phys.\ Lett.\ B {\bf 783}, 13 (2018)
  %doi:10.1016/j.physletb.2018.06.039
  [arXiv:1710.01743 [hep-ph]].
  %%CITATION = doi:10.1016/j.physletb.2018.06.039;%%
  %20 citations counted in INSPIRE as of 07 Nov 2018 
 
 %\cite{CidVidal:2018blh}
\bibitem{CidVidal:2018blh} 
  X.~C.~Vidal, A.~Mariotti, D.~Redigolo, F.~Sala and K.~Tobioka,
  %``New Axion Searches at Flavor Factories,''
  arXiv:1810.09452 [hep-ph].
  %%CITATION = ARXIV:1810.09452;%%
 
 
 %\cite{Bochkarev:1987wf,Cline:2000nw}
\bibitem{Bochkarev:1987wf} 
  A.~I.~Bochkarev and M.~E.~Shaposhnikov,
  %``Electroweak Production of Baryon Asymmetry and Upper Bounds on the Higgs and Top Masses,''
  Mod.\ Phys.\ Lett.\ A {\bf 2}, 417 (1987).
%  doi:10.1142/S0217732387000537
  %%CITATION = doi:10.1142/S0217732387000537;%%
  %190 citations counted in INSPIRE as of 06 Nov 2018
 
 %\cite{Cline:2000nw}
\bibitem{Cline:2000nw} 
  J.~M.~Cline, M.~Joyce and K.~Kainulainen,
  %``Supersymmetric electroweak baryogenesis,''
  JHEP {\bf 0007}, 018 (2000)
%  doi:10.1088/1126-6708/2000/07/018
  [hep-ph/0006119].
  %%CITATION = doi:10.1088/1126-6708/2000/07/018;%%
  %202 citations counted in INSPIRE as of 06 Nov 2018
  
  
  %\cite{Cohen:1991iu,Giudice:1993bb,Dine:1994vf,Joyce:1994bk,Cohen:1994ss,Joyce:1994fu}
\bibitem{Cohen:1991iu} 
  A.~G.~Cohen, D.~B.~Kaplan and A.~E.~Nelson,
  %``Spontaneous baryogenesis at the weak phase transition,''
  Phys.\ Lett.\ B {\bf 263}, 86 (1991).
%  doi:10.1016/0370-2693(91)91711-4
  %%CITATION = doi:10.1016/0370-2693(91)91711-4;%%
  %254 citations counted in INSPIRE as of 06 Nov 2018
  
  %\cite{Giudice:1993bb,Dine:1994vf,Joyce:1994bk,Cohen:1994ss,Joyce:1994fu}
\bibitem{Giudice:1993bb} 
  G.~F.~Giudice and M.~E.~Shaposhnikov,
  %``Strong sphalerons and electroweak baryogenesis,''
  Phys.\ Lett.\ B {\bf 326}, 118 (1994)
 % doi:10.1016/0370-2693(94)91202-5
  [hep-ph/9311367].
  %%CITATION = doi:10.1016/0370-2693(94)91202-5;%%
  %79 citations counted in INSPIRE as of 06 Nov 2018
  
  %\cite{Dine:1994vf,Joyce:1994bk,Cohen:1994ss,Joyce:1994fu}
\bibitem{Dine:1994vf} 
  M.~Dine and S.~D.~Thomas,
  %``Electroweak baryogenesis in the adiabatic limit,''
  Phys.\ Lett.\ B {\bf 328}, 73 (1994)
 % doi:10.1016/0370-2693(94)90430-8
  [hep-ph/9401265].
  %%CITATION = doi:10.1016/0370-2693(94)90430-8;%%
  %38 citations counted in INSPIRE as of 06 Nov 2018
  
  %\cite{Joyce:1994bk,Cohen:1994ss,Joyce:1994fu}
\bibitem{Joyce:1994bk} 
  M.~Joyce, T.~Prokopec and N.~Turok,
  %``Constraints and transport in electroweak baryogenesis,''
  Phys.\ Lett.\ B {\bf 339}, 312 (1994)
  [hep-ph/9401351].
  %%CITATION = HEP-PH/9401351;%%
  %34 citations counted in INSPIRE as of 06 Nov 2018
  
  %\cite{Cohen:1994ss,Joyce:1994fu}
\bibitem{Cohen:1994ss} 
  A.~G.~Cohen, D.~B.~Kaplan and A.~E.~Nelson,
  %``Diffusion enhances spontaneous electroweak baryogenesis,''
  Phys.\ Lett.\ B {\bf 336}, 41 (1994)
  %doi:10.1016/0370-2693(94)00935-X
  [hep-ph/9406345].
  %%CITATION = doi:10.1016/0370-2693(94)00935-X;%%
  %110 citations counted in INSPIRE as of 06 Nov 2018
  
  %\cite{Joyce:1994fu}
\bibitem{Joyce:1994fu} 
  M.~Joyce, T.~Prokopec and N.~Turok,
  %``Electroweak baryogenesis from a classical force,''
  Phys.\ Rev.\ Lett.\  {\bf 75}, 1695 (1995)
  Erratum: [Phys.\ Rev.\ Lett.\  {\bf 75}, 3375 (1995)]
  %doi:10.1103/PhysRevLett.75.3375, 10.1103/PhysRevLett.75.1695
  [hep-ph/9408339].
  %%CITATION = doi:10.1103/PhysRevLett.75.3375, 10.1103/PhysRevLett.75.1695;%%
  %111 citations counted in INSPIRE as of 06 Nov 2018
 
 
 %\cite{DOnofrio:2014rug}
\bibitem{DOnofrio:2014rug} 
  M.~D'Onofrio, K.~Rummukainen and A.~Tranberg,
  %``Sphaleron Rate in the Minimal Standard Model,''
  Phys.\ Rev.\ Lett.\  {\bf 113}, no. 14, 141602 (2014)
  %doi:10.1103/PhysRevLett.113.141602
  [arXiv:1404.3565 [hep-ph]].
  %%CITATION = doi:10.1103/PhysRevLett.113.141602;%%
  %101 citations counted in INSPIRE as of 06 Nov 2018
 
 
 
 %\cite{Shaposhnikov:1987pf,Dine:1990fj}
\bibitem{Shaposhnikov:1987pf} 
  M.~E.~Shaposhnikov,
  %``Structure of the High Temperature Gauge Ground State and Electroweak Production of the Baryon Asymmetry,''
  Nucl.\ Phys.\ B {\bf 299}, 797 (1988).
  %doi:10.1016/0550-3213(88)90373-2
  %%CITATION = doi:10.1016/0550-3213(88)90373-2;%%
  %347 citations counted in INSPIRE as of 06 Nov 2018
  
  %\cite{Dine:1990fj}
\bibitem{Dine:1990fj} 
  M.~Dine, P.~Huet, R.~L.~Singleton, Jr and L.~Susskind,
  %``Creating the baryon asymmetry at the electroweak phase transition,''
  Phys.\ Lett.\ B {\bf 257}, 351 (1991).
  %doi:10.1016/0370-2693(91)91905-B
  %%CITATION = doi:10.1016/0370-2693(91)91905-B;%%
  %218 citations counted in INSPIRE as of 06 Nov 2018
 
 
 
 %\cite{GarciaBellido:1999sv,Konstandin:2011ds,Servant:2014bla}
\bibitem{GarciaBellido:1999sv} 
  J.~Garcia-Bellido, D.~Y.~Grigoriev, A.~Kusenko and M.~E.~Shaposhnikov,
  %``Nonequilibrium electroweak baryogenesis from preheating after inflation,''
  Phys.\ Rev.\ D {\bf 60}, 123504 (1999)
  %doi:10.1103/PhysRevD.60.123504
  [hep-ph/9902449].
  %%CITATION = doi:10.1103/PhysRevD.60.123504;%%
  %205 citations counted in INSPIRE as of 06 Nov 2018
 
 
 
 %\cite{Konstandin:2011ds,Servant:2014bla}
\bibitem{Konstandin:2011ds} 
  T.~Konstandin and G.~Servant,
  %``Natural Cold Baryogenesis from Strongly Interacting Electroweak Symmetry Breaking,''
  JCAP {\bf 1107}, 024 (2011)
 % doi:10.1088/1475-7516/2011/07/024
  [arXiv:1104.4793 [hep-ph]].
  %%CITATION = doi:10.1088/1475-7516/2011/07/024;%%
  %30 citations counted in INSPIRE as of 06 Nov 2018
 
 %\cite{Servant:2014bla}
\bibitem{Servant:2014bla} 
  G.~Servant,
  %``Baryogenesis from Strong $CP$ Violation and the QCD Axion,''
  Phys.\ Rev.\ Lett.\  {\bf 113}, no. 17, 171803 (2014)
  %doi:10.1103/PhysRevLett.113.171803
  [arXiv:1407.0030 [hep-ph]].
  %%CITATION = doi:10.1103/PhysRevLett.113.171803;%%
  %27 citations counted in INSPIRE as of 06 Nov 2018
 
 
 %\cite{Craig:2010au}
\bibitem{Craig:2010au} 
N.~Craig and J.~March-Russell,
%``Axion-Assisted Electroweak Baryogenesis,''
arXiv:1007.0019 [hep-ph].
%%CITATION = ARXIV:1007.0019;%%
%5 citations counted in INSPIRE as of 09 Sep 2019 
 
 %\cite{Graham:2015cka}
\bibitem{Graham:2015cka} 
  P.~W.~Graham, D.~E.~Kaplan and S.~Rajendran,
  %``Cosmological Relaxation of the Electroweak Scale,''
  Phys.\ Rev.\ Lett.\  {\bf 115}, no. 22, 221801 (2015)
  %doi:10.1103/PhysRevLett.115.221801
  [arXiv:1504.07551 [hep-ph]].
  %%CITATION = doi:10.1103/PhysRevLett.115.221801;%%
  %233 citations counted in INSPIRE as of 06 Nov 2018
  
  
  %\cite{Yokoyama:1998ju}
%\bibitem{Yokoyama:1998ju} 
%  J.~Yokoyama and A.~D.~Linde,
  %``Is warm inflation possible?,''
%  Phys.\ Rev.\ D {\bf 60}, 083509 (1999)
  %doi:10.1103/PhysRevD.60.083509
%  [hep-ph/9809409].
  %%CITATION = doi:10.1103/PhysRevD.60.083509;%%
  %113 citations counted in INSPIRE as of 06 Nov 2018
  
 %\cite{Wang:1999mb}
\bibitem{Wang:1999mb} 
  S.~Y.~Wang, D.~Boyanovsky, H.~J.~de Vega, D.~S.~Lee and Y.~J.~Ng,
  %``Damping rates and mean free paths of soft fermion collective excitations in a hot fermion gauge scalar theory,''
  Phys.\ Rev.\ D {\bf 61}, 065004 (2000)
  doi:10.1103/PhysRevD.61.065004
  [hep-ph/9902218].
  %%CITATION = doi:10.1103/PhysRevD.61.065004;%%
  %23 citations counted in INSPIRE as of 11 Dec 2019 
 
 %\cite{Mukaida:2012qn}
\bibitem{Mukaida:2012qn} 
  K.~Mukaida and K.~Nakayama,
  %``Dynamics of oscillating scalar field in thermal environment,''
  JCAP {\bf 1301}, 017 (2013)
  doi:10.1088/1475-7516/2013/01/017
  [arXiv:1208.3399 [hep-ph]].
  %%CITATION = doi:10.1088/1475-7516/2013/01/017;%%
  %66 citations counted in INSPIRE as of 11 Dec 2019
 
 
  
  %\cite{Choi:2016luu}
\bibitem{Choi:2016luu} 
  K.~Choi and S.~H.~Im,
  %``Constraints on Relaxion Windows,''
  JHEP {\bf 1612}, 093 (2016)
  %doi:10.1007/JHEP12(2016)093
  [arXiv:1610.00680 [hep-ph]].
  %%CITATION = doi:10.1007/JHEP12(2016)093;%%
  %28 citations counted in INSPIRE as of 06 Nov 2018

%\cite{Flacke:2016szy}
\bibitem{Flacke:2016szy} 
  T.~Flacke, C.~Frugiuele, E.~Fuchs, R.~S.~Gupta and G.~Perez,
  %``Phenomenology of relaxion-Higgs mixing,''
  JHEP {\bf 1706}, 050 (2017)
  %doi:10.1007/JHEP06(2017)050
  [arXiv:1610.02025 [hep-ph]].
  %%CITATION = doi:10.1007/JHEP06(2017)050;%%
  %51 citations counted in INSPIRE as of 06 Nov 2018
  
  
  
  
  
  
    %\cite{Aaij:2012vr,Wei:2009zv,Aaij:2015tna, Bezrukov:2014nza, Hyun:2010an}
\bibitem{Aaij:2012vr} 
  R.~Aaij {\it et al.} [LHCb Collaboration],
  %``Differential branching fraction and angular analysis of the $B^{+} \rightarrow K^{+}\mu^{+}\mu^{-}$ decay,''
  JHEP {\bf 1302}, 105 (2013)
  %doi:10.1007/JHEP02(2013)105
  [arXiv:1209.4284 [hep-ex]].
  %%CITATION = doi:10.1007/JHEP02(2013)105;%%
  %95 citations counted in INSPIRE as of 10 Oct 2018

%\cite{Wei:2009zv,Aaij:2015tna, Bezrukov:2014nza}
\bibitem{Wei:2009zv} 
  J.-T.~Wei {\it et al.} [Belle Collaboration],
  %``Measurement of the Differential Branching Fraction and Forward-Backword Asymmetry for $B \to K^{(*)}\ell^+\ell^-$,''
  Phys.\ Rev.\ Lett.\  {\bf 103}, 171801 (2009)
  %doi:10.1103/PhysRevLett.103.171801
  [arXiv:0904.0770 [hep-ex]].
  %%CITATION = doi:10.1103/PhysRevLett.103.171801;%%
  %418 citations counted in INSPIRE as of 10 Oct 2018

%\cite{Aaij:2015tna, Bezrukov:2014nza}
\bibitem{Aaij:2015tna} 
  R.~Aaij {\it et al.} [LHCb Collaboration],
  %``Search for hidden-sector bosons in $B^0 \!\to K^{*0}\mu^+\mu^-$ decays,''
  Phys.\ Rev.\ Lett.\  {\bf 115}, no. 16, 161802 (2015)
 % doi:10.1103/PhysRevLett.115.161802
  [arXiv:1508.04094 [hep-ex]].
  %%CITATION = doi:10.1103/PhysRevLett.115.161802;%%
  %58 citations counted in INSPIRE as of 10 Oct 2018

%\cite{Bezrukov:2014nza}
\bibitem{Bezrukov:2014nza} 
  F.~Bezrukov and D.~Gorbunov,
  %``Relic Gravity Waves and 7 keV Dark Matter from a GeV scale inflaton,''
  Phys.\ Lett.\ B {\bf 736}, 494 (2014)
 % doi:10.1016/j.physletb.2014.07.060
  [arXiv:1403.4638 [hep-ph]].
  %%CITATION = doi:10.1016/j.physletb.2014.07.060;%%
  %63 citations counted in INSPIRE as of 10 Oct 2018

%\cite{Hyun:2010an}
\bibitem{Hyun:2010an} 
  H.~J.~Hyun {\it et al.} [Belle Collaboration],
  %``Search for a Low Mass Particle Decaying into mu^+ mu^- in B^0 -> K^{*0} X and B^0 -> rho^0 X at Belle,''
  Phys.\ Rev.\ Lett.\  {\bf 105}, 091801 (2010)
%  doi:10.1103/PhysRevLett.105.091801
  [arXiv:1005.1450 [hep-ex]].
  %%CITATION = doi:10.1103/PhysRevLett.105.091801;%%
  %26 citations counted in INSPIRE as of 10 Oct 2018
  
  
  
%\cite{Adler:2004hp}\cite{Artamonov:2009sz}
\bibitem{Adler:2004hp} 
  S.~Adler {\it et al.} [E787 Collaboration],
  %``Further search for the decay K+ ---> pi+ nu anti-nu in the momentum region P < 195-MeV/c,''
  Phys.\ Rev.\ D {\bf 70}, 037102 (2004)
%  doi:10.1103/PhysRevD.70.037102
  [hep-ex/0403034].
  %%CITATION = doi:10.1103/PhysRevD.70.037102;%%
  %88 citations counted in INSPIRE as of 10 Oct 2018


%\cite{Artamonov:2009sz}
\bibitem{Artamonov:2009sz}
  A.~V.~Artamonov {\it et al.} [BNL-E949 Collaboration],
  %``Study of the decay $K^+\to\pi^+\nu \bar\nu$ in the momentum region $140 < P_\pi < 199$ MeV/c,''
  Phys.\ Rev.\ D {\bf 79} (2009) 092004
  %doi:10.1103/PhysRevD.79.092004
  [arXiv:0903.0030 [hep-ex]].
  %%CITATION = doi:10.1103/PhysRevD.79.092004;%%
  %239 citations counted in INSPIRE as of 10 Oct 2018
  
  
  %\cite{Bergsma:1985qz}
\bibitem{Bergsma:1985qz} 
  F.~Bergsma {\it et al.} [CHARM Collaboration],
  %``Search for Axion Like Particle Production in 400-{GeV} Proton - Copper Interactions,''
  Phys.\ Lett.\  {\bf 157B}, 458 (1985).
  %doi:10.1016/0370-2693(85)90400-9
  %%CITATION = doi:10.1016/0370-2693(85)90400-9;%%
  %106 citations counted in INSPIRE as of 10 Oct 2018
  
  %\cite{Alekhin:2015byh}
\bibitem{Alekhin:2015byh} 
  S.~Alekhin {\it et al.},
  %``A facility to Search for Hidden Particles at the CERN SPS: the SHiP physics case,''
  Rept.\ Prog.\ Phys.\  {\bf 79}, no. 12, 124201 (2016)
  %doi:10.1088/0034-4885/79/12/124201
  [arXiv:1504.04855 [hep-ph]].
  %%CITATION = doi:10.1088/0034-4885/79/12/124201;%%
  %295 citations counted in INSPIRE as of 10 Oct 2018


\bibitem{McLerran:1990de} 
  L.~D.~McLerran, E.~Mottola and M.~E.~Shaposhnikov,
  %``Sphalerons and Axion Dynamics in High Temperature {QCD},''
  Phys.\ Rev.\ D {\bf 43}, 2027 (1991).
  %doi:10.1103/PhysRevD.43.2027
  %%CITATION = doi:10.1103/PhysRevD.43.2027;%%
  %109 citations counted in INSPIRE as of 08 Sep 2019

%\cite{Burnier:2005hp}
\bibitem{Burnier:2005hp} 
  Y.~Burnier, M.~Laine and M.~Shaposhnikov,
  %``Baryon and lepton number violation rates across the electroweak crossover,''
  JCAP {\bf 0602}, 007 (2006)
  %doi:10.1088/1475-7516/2006/02/007
  [hep-ph/0511246].
  %%CITATION = doi:10.1088/1475-7516/2006/02/007;%%
  %79 citations counted in INSPIRE as of 08 Sep 2019


  \end{thebibliography}
 \end{document}